\pdfoutput=1
\documentclass[%
 reprint,
 amsmath,amssymb,
 aps,
]{revtex4-1}

\usepackage{graphicx}
\usepackage{dcolumn}
\usepackage{bm}
\usepackage{hyperref}
\hypersetup{urlcolor=blue,linkcolor=black,citecolor=black,colorlinks=true}

\begin{document}

\preprint{APS/123-QED}

\title{Clocked Atom Delivery to a Photonic Crystal Waveguide}

\author{A. P. Burgers}
 \email{aburgers@caltech.edu}
\author{L. S. Peng}
\author{J. A. Muniz}
 \altaffiliation[Present address: ]{JILA, University of Colorado, Boulder, CO 80309.}
\author{A. C. McClung}%
 \altaffiliation[Present address: ]{Department of Electrical and Computer Engineering, University of Massachusetts Amherst, 151 Holdsworth Way, Amherst, MA 01003.}
\author{M. J. Martin}%
 \altaffiliation[Present address: ]{P-21, Physics Division, Los Alamos National Laboratory, Los Alamos, NM 87545.}
\author{H. J. Kimble}
 \email{hjkimble@caltech.edu}
\affiliation{Norman Bridge Laboratory of Physics MC12-33, California Institute of Technology, Pasadena, CA 91125.}


\date{\today}

\begin{abstract}
 Experiments and numerical simulations are described that develop quantitative understanding of atomic motion near the surfaces of nanoscopic photonic crystal waveguides (PCWs). Ultracold atoms are delivered from a moving optical lattice into the PCW. Synchronous with the moving lattice, transmission spectra for a guided-mode probe field are recorded as functions of lattice transport time and frequency detuning of the probe beam. By way of measurements such as these, we have been able to validate quantitatively our numerical simulations, which are based upon detailed understanding of atomic trajectories that pass around and through nanoscopic regions of the PCW under the influence of optical and surface forces. The resolution for mapping atomic motion is roughly 50 nm in space and 100 ns in time. By introducing auxiliary guided mode (GM) fields that provide spatially varying AC-Stark shifts, we have, to some degree, begun to control atomic trajectories, such as to enhance the flux into to the central vacuum gap of the PCW at predetermined times and with known AC-Stark shifts. Applications of these capabilities include enabling high fractional filling of optical trap sites within PCWs, calibration of optical fields within PCWs, and utilization of the time-dependent, optically dense atomic medium for novel nonlinear optical experiments.
\end{abstract}

\pacs{Valid PACS appear here}
\maketitle


\section{Introduction}

The integration of ultra-cold atoms with nanophotonic devices leads to strong atom-photon interactions that not only quantitatively advance Quantum Optics but which also create qualitatively new paradigms for atom-photon interactions \cite{chang2018colloquium,lodahl2017chiral,pichler2017universal}. Owing to small optical loss and tight confinement of the electromagnetic field, nanoscale devices are capable of mediating long-range atom-atom interactions by way of optical photons in the guided modes (GMs) of the structures. In a complimentary fashion, long-range interactions between photons can be mediated by underlying lattices of atoms in one and two-dimensional dielectric structures \cite{gonzalez2015subwavelength,hung2016quantum,gonzalez2017quantum}. With advanced fabrication capabilities brought by nanophotonics, such physical processes provide novel possibilities for quantum communication, computation, networks, and metrology, as well as for quantum phases of light and matter \cite{chang2018colloquium,lodahl2017chiral,pichler2017universal,gonzalez2015subwavelength,hung2016quantum,gonzalez2017quantum}.

Compared to the rapidly expanding theoretical literature, to date experimental progress has been rather more modest in integrating atoms and nanophotonics. Important laboratory systems include optical nanofibers, where $\simeq10^3$ atoms have been trapped $\simeq 220$ nm from fiber surfaces \cite{vetsch2010optical,Goban2012Demonstration,beguin2014generation,gouraud2015demonstration,solano2017optical}. Dispersion engineered photonic crystal cavities \cite{thompson2013coupling,tiecke2014nanophotonic} and waveguides \cite{gobanprl,hood2016atom} have achieved strong atom-photon coupling, albeit with only a few atoms trapped $\simeq 150$ nm from dielectric surfaces. A grand challenge for this emerging field remains the laboratory attainment of one and two-dimensional atomic lattices with high filling fraction and strong coupling of single atoms to single photons within the GMs of the nanophotonic structures.
 
For the photonic crystal waveguides [PCWs] considered here, strong atom-field coupling requires devices designed for atomic physics that provide both stable atom trapping and large atom-photon coupling at the atom trapping sites \cite{chang2018colloquium,hung2013trapped}. In the optical domain, suitable PCWs have lattice constant $a \simeq 350$ nm for dielectric constant $\epsilon \simeq 4$. Single-atom localization with optical traps inside vacuum voids of unit cells then constrains far-off resonance traps (FORTs \cite{grimm2000optical}) to volume $(\Delta x, \Delta y, \Delta z) = (30, 100, 140)$ nm for energy of 100 $\mu$K where the coordinate system for the structure is given in Fig. \ref{fig:sem}. Free-space atoms must be transported to and cooled within these tiny FORTs. Such transport, cooling, and trapping of atoms near and within nanoscopic dielectric structures requires adaptations of existing techniques from atomic physics \cite{dalibard1989laser,wieman1999atom,phillips1998nobel,bannerman2009single}, as well as the invention of new protocols, such as hybrid vacuum-light traps \cite{chang2018colloquium,hung2013trapped}.
 
\begin{figure}[t]
\centering
\includegraphics[width=1\linewidth]{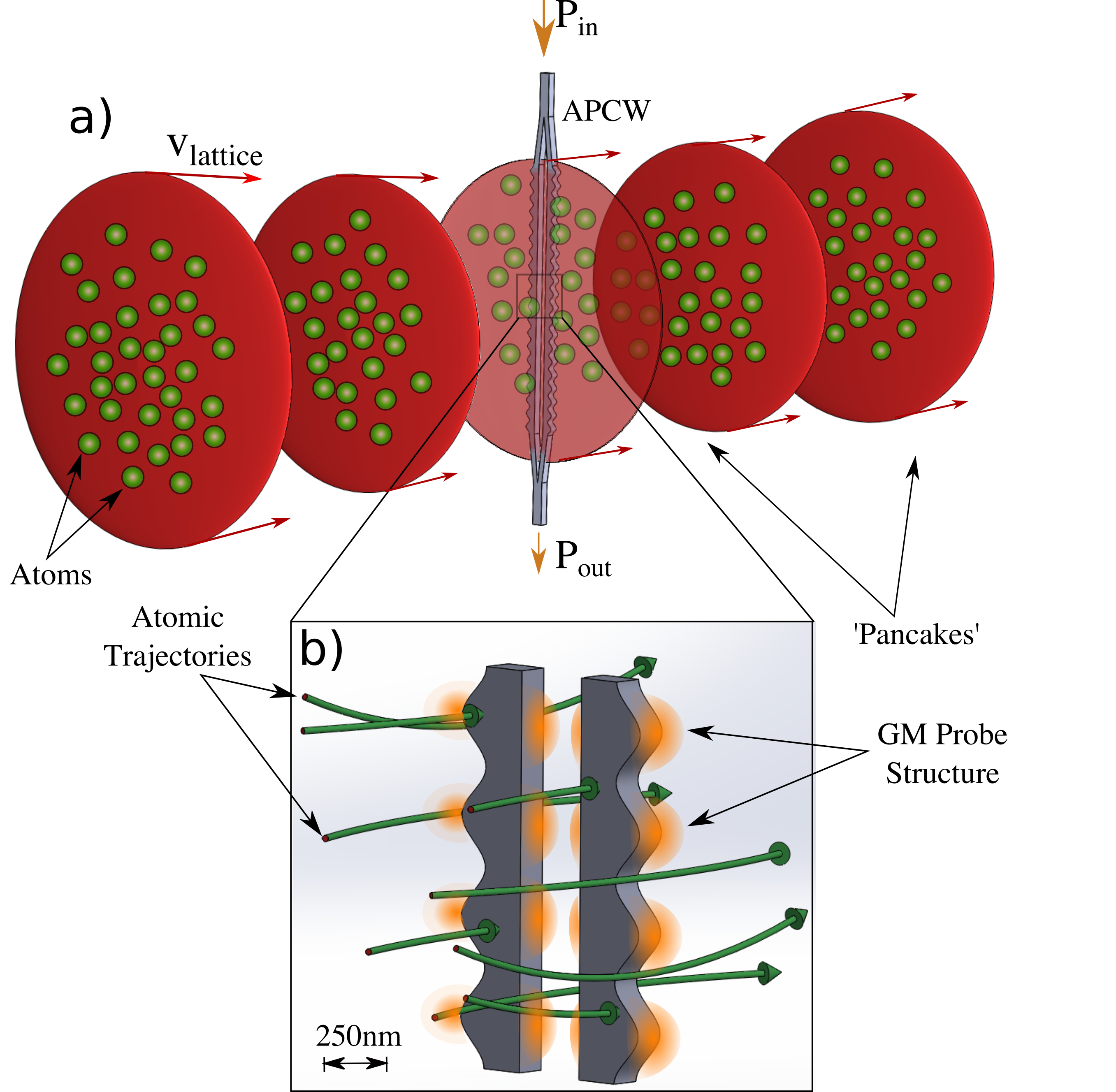}
\caption{(a) Atomic delivery to the `alligator' photonic crystal waveguide (APCW) in our experiment is achieved by creating an optical conveyor belt from a 1D free-space lattice. The atoms, trapped in `pancakes' of high laser intensity are transported to the device at a speed $v_{\text{lattice}}$ and interact transiently with the APCW. The atomic signature is read out through changes in the transmission of a weak, near resonant probe (P$_{\text{in}}\rightarrow\text{P}_{\text{out}}$) that is an electromagnetic mode of the waveguide. (b) Depiction of atom trajectories showing the transient nature of the atoms as they pass through, around and `crash' into the APCW. The bright probe modes on the `teeth' and center of APCW illustrate the TE-like guided mode (GM) intensity near a TE bandedge (here the dielectric bandedge at $\simeq894$ nm (D1 line of Cesium)). While these trajectories are merely an illustration, calculated atomic trajectories are presented along with comparisons to experimental data in Fig. \ref{fig:tmtmsimdat}.}\label{fig:setup}
\end{figure}

Thereby motivated, in this article we describe various investigations aimed at developing better quantitative understanding and new tools for the control of atomic motion under the influence of optical and surface forces near nanophotonic PCWs, as illustrated in Fig. \ref{fig:setup}. An important goal of this research is to formulate and validate \textit{in situ} diagnostics that enable atoms to be conveyed from free space into GM optical traps within a PCW, ultimately with high fractional filling of each lattice site.
 
More specifically, our system shown in Fig. \ref{fig:setup} consists of a quasi-one-dimensional PCW whose band structure arises from periodic modulation of the outer edges of two parallel dielectric beams with a central vacuum gap, resulting in the so-called `Alligator Photonic Crystal Waveguide' (or `APCW') \cite{yu2014nanowire,gobanprl,hood2016atom}. A moving optical lattice transports trapped atoms into and through the APCW with velocity $v_{\text{lattice}} \simeq0.5$ m/s and temperature $T_{\text{lattice}}\simeq20 \mu$K. Synchronous with the moving lattice, transmission spectra are recorded as functions of lattice transport time and detuning of weak GM probes. Due to the periodicity of the lattice delivery, the recorded spectra can be offset in time by integer multiples of the lattice period and coherently combined to create time-dependent `clocked spectra' with temporal (spatial) resolution of $\simeq 100$ ns ($\simeq 50$ nm). 
 
These measurements allow us to quantitatively validate our numerical simulations and there by provide understanding of atomic trajectories passing around and through nanoscopic regions of the APCW. For example, we infer that the number of atoms transported into the 250 nm wide vacuum gap of the APCW during each lattice period is of order unity (i.e., $\simeq10^6$ atoms/sec into the central vacuum gap). Introducing auxiliary GM fields provides a spatially varying AC-Stark shift, which can control atomic trajectories arriving to the central vacuum gap at predetermined times and with known AC-Stark shifts. 
 
This quantitative understanding of atomic transport through the nanoscopic APCW provides new tools for the integration of ultracold atoms and nanophotonics, some of which we describe here. For example, following Ref. \cite{bannerman2009single}, we consider the transfer of atoms from the moving optical lattice into GM trap sites along the APCW by way of single-photon scattering events triggered by the temporal phase of the moving lattice.
 
The research reported in this paper has many important antecedents in AMO physics. Historic measurements of transmission for atomic beams passing through nano-fabricated arrays of slits led to among the first observations of interferometry with atomic deBroglie waves, that were followed over the next decades by precision measurements of many fundamental atomic properties (Ref. \cite{cronin2009optics} for a review), including atom-surface interactions important to our work. Pioneering experiments to measure atomic line shifts and decay rate modifications for atoms near surfaces were also made in other microscopic geometries with atomic transmission recorded through various structures \cite{haroche1989cavity,sukenik1993measurement,intravaia2011fluctuation}, which are likewise quite relevant for our system. Early work in atomic vapors lead to  measurements of the spectral line distortion for Cs atoms induced by surface forces on the atoms \cite{oria1991spectral,chevrollier1991van,bloch2005atom}. More recent experiments investigate the role of Casimir-Polder (CP) forces on simulated trajectories of thermal ($\simeq300$ K) atoms interacting with nanoscopic slot waveguides and accompanying experiments verifying the importance of CP when explaining the experimental result  \cite{ritter2018coupling}.
Among many experiments with laser cooled atoms, landmark measurements of CP forces were made by `bouncing' atoms from evanescent fields \cite{landragin1996measurement}, as well by utilizing interactions with BECs at controlled distances from a dielectric boundary, again with atom loss being the relevant variable \cite{lin2004impact,obrecht2007measurement}. Perhaps closest in spirit and implementation to the present work are pioneering experiments and numerical simulations for cold atoms moving near nanoscopic optical fibers \cite{sague2007cold}, micro-torodial resonators \cite{stern2011simulations,alton2011strong}, and PCWs \cite{gobanncomm} in which light transmission and reflection from the respective optical structures were employed to link experiment and numerical simulation of atomic motion, rather than by direct atomic detection.

\section{Photonic Crystal Waveguide and Supporting Structure}

This section provides an overview of the APCW utilized in our experiments, including the relevant GMs of the waveguide, the coupling of light into and out of these GMs and the structural support of the waveguide. More details about device fabrication and characterization can be found in Refs. \cite{yu2014nanowire,yu2017nano,mcclung2017photonic}.
 
The APCWs used in our experiments are made by pattering the desired geometry using electron beam lithography on a $200$ nm-thick layer of Silicon Nitride (SiN) grown on a $200$ $\mu$m Silicon substrate, followed by various stages of chemical processing \cite{mcclung2017photonic,yu2017nano}. The photonic crystal is formed by external sinusoidal modulation of two parallel nano-beams to create a photonic bandgap for TE modes with band edges at the strong dipole-allowed transition lines of Cesium (Cs), the D1 line near 894 nm and the D2 line near 852 nm.
 
\begin{figure}[t]
\centering
\includegraphics[width=1\linewidth]{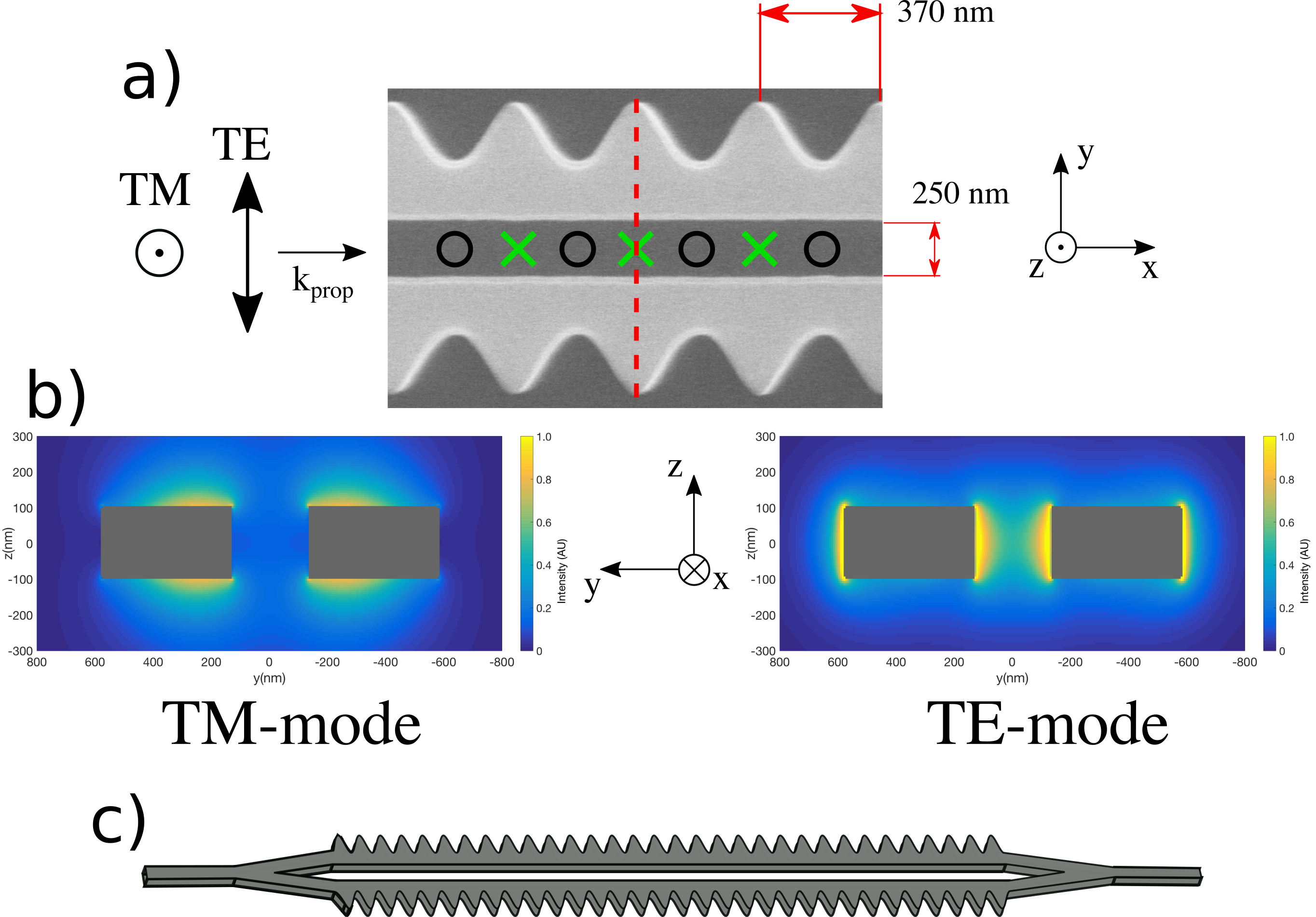}
\caption{Details of the alligator photonic crystal waveguide (APCW) used in our experiments. a) An SEM image of 4 unit cells of the APCW taken in the center region of the waveguide.  The unit cell spacing is 370 nm, the vacuum gap between the beams is 250 nm, and the Silicon Nitride is 200 nm thick. `X's (`O's) indicate the regions of the APCW where TE guided light at the dielectric or D1 (air or D2) bandegde is brighteste. b) The intensity distribution of the TE and TM polarizations supported by the structure at a single slice in the $y$ direction, indicated by the red dashed line in (a). c) The 150 unit cells of the APCW are formed from single rectangular waveguides on either end that split at Y-junctions into parallel, modulated beams. The entire structure is suspended in vacuum by transverse tethers connected to supporting side rails (not shown) \cite{yu2014nanowire,yu2017nano,mcclung2017photonic}.}\label{fig:sem}
\end{figure}

Fig. \ref{fig:sem}(a) displays a scanning-electron microscope (SEM) image of the central section of an APCW. The `X' and `O' points indicate regions of high GM intensity for TE-polarized light with frequencies near the dielectric (D1) and air (D2) bandedges, respectively. The coordinate system adopted for our subsequent analyses is shown together with the dimensions of the device and the principal polarizations of relevant GMs supported by the dielectric structure. Calculated and measured dispersion relations for such devices are presented in Ref \cite{hood2016atom} where good quantitative agreement is found. 
 
Fig. \ref{fig:sem}(b) provides a cross-section of the two dielectric beams that form the PCW overlaid with intensity profiles for the TM-like and TE-like polarizations for the $x$ coordinate corresponding to the widest sections of the APCW. Note that the TM mode's regions of highest intensity are on the top and bottom surfaces of the dielectric, while the TE mode is primarily bright in the center (vacuum gap between the beams) and on the sides. For operation near the band edges of the TE mode, spontaneous emission rates, $\Gamma_{\text{1D}}$, into the waveguide are strongest for atoms coupled to TE GMs ($\Gamma_{\text{1D}}/\Gamma^{\prime}\leqslant10$), and significantly weaker for emission into TM GMs \cite{yu2014nanowire,hood2016atom}. Additionally, near the TE bandedges for D1 and D2, the Bloch functions develop high-contrast standing waves along $x$, which are useful for creating dipole trapping potentials (FORTs) for atoms within the waveguide \cite{hung2013trapped,yu2014nanowire}.  The TM GMs, near D1 and D2 have low contrast in the $x$ direction and resemble simple GMs of an unstructured waveguide. A figure of these modes and the resulting trapping potentials is provided in SI V. 

The structure depicted in Fig. \ref{fig:sem}(c) illustrates the APCW connected to single-beam waveguides on either end and thereby suspended in the center of a  2 mm wide window in the Silicon chip to allow optical access for delivering and manipulating cold atoms near the APCW. Well beyond the ends of the APCW, a series of tethers are attached transversely to the single-beam waveguides along $\pm y$ to anchor the waveguides to two side rails that run parallel to the $x$ axis of the device to provide thermal anchoring and mechanical support.

\begin{figure*}[t]
\centering
\includegraphics[width=\textwidth]{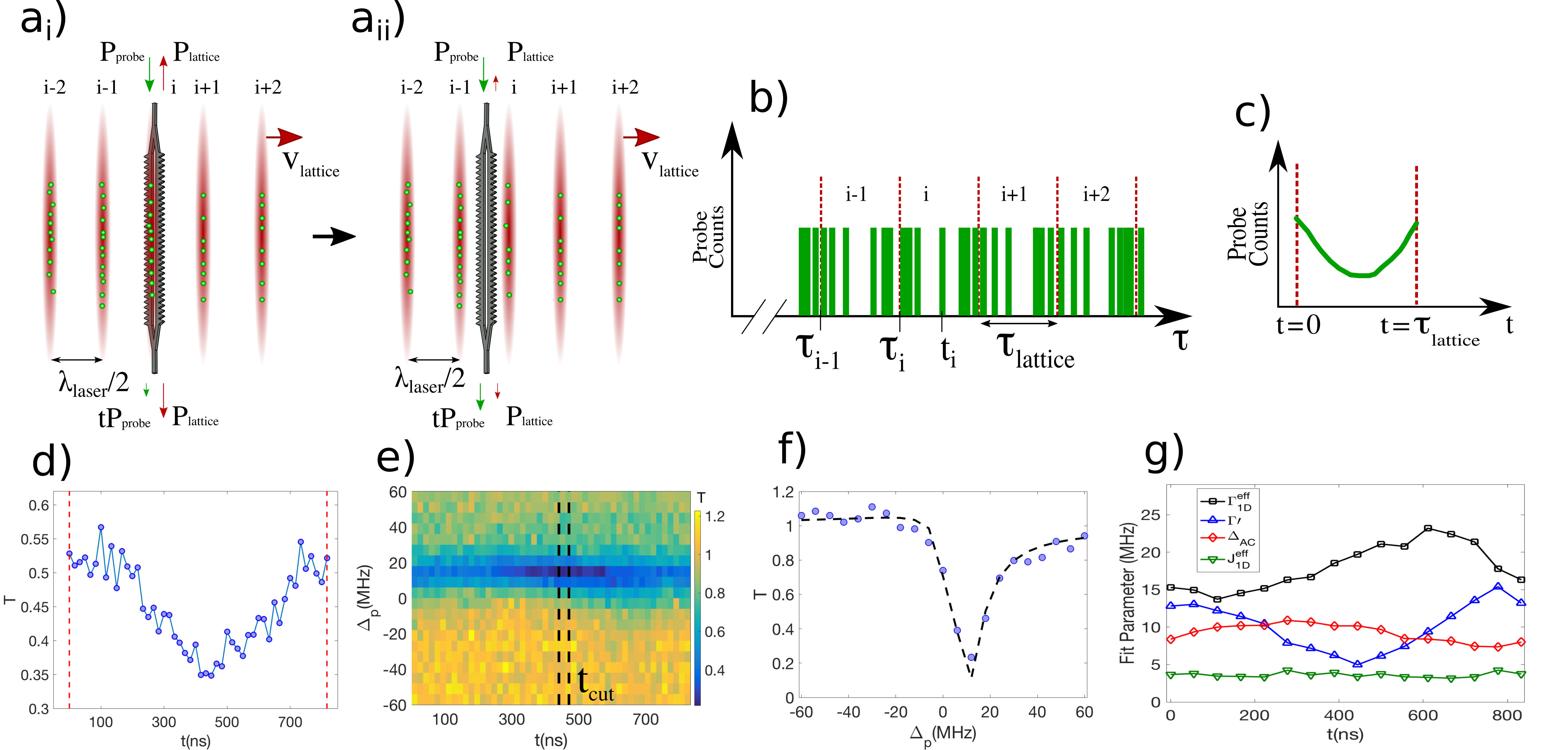}
\caption{This figure details the concept of the clocked atom delivery and how the data is analyzed to recover a phase sensitive atomic signal. ai) Atoms trapped in pancakes by the 1D lattice are delivered to the structure when a chirp sequence is placed on one of the lattice beams creating a conveyor belt.  aii) A half period later the `pancakes' have moved $\lambda_\text{{laser}}/4$ relative to the APCW. b) Probe photon counts placed on the same time axis as the lattice sync signal. By building a histogram from probe time stamps ($t_i$) between successive lattice sync time stamps ($\tau_i,\tau_{i-1}...$) we can recover the phase sensitive nature of the atom delivery. c) All the individual lattice time periods are folded together into a single histogram indicating that the atomic signature is phased with the lattice signature. d) For a single probe detuning we can observe this clear phase sensitive signal when the probe photon time stamps are referenced to the closest time stamps of the lattice.  e) By scanning the probe detuning we can create this 2D image in lattice time $t$ and probe detuning $\Delta_\text{p}$ to gain more insight into the arrival time of different classes of atoms and extract spectra for various arrival times. This information helps to determine the coupling of the atoms to the waveguide and can reveal time dependent AC-Stark shifts. Cuts at certain times reveal spectra which we fit to extract parameters $\{\Gamma^{\text{eff}}_{\text{1D}}(t),\Gamma^{\prime}(t),\Delta_{\text{AC}}(t),J^{\text{eff}}_{\text{1D}}(t)\}$ for an effective model. An example of slices taken over an interval `$t_{\text{cut}}$' and summed is given in (f). g) The fit parameters extracted for spectra at  varying lattice times $t$ in (e). Each spectrum is taken by combining 3 time bins ($\simeq50$ ns) and fitting to a transmission model given in Ref. \cite{hood2016atom} and the SI II. Note that $\Gamma^{\text{eff}}_{\text{1D}}$ changes the most across time and denotes the coupling strength of the atomic sample to the structure. $J^{\text{eff}}_{\text{1D}}$, the coherent coupling term, remains relatively constant and small across time.}\label{fig:clock} 
\end{figure*}

Light is coupled into and out of the GMs of the APCW by conventional optical fibers that are mode-matched to the fields to/from the terminating ends of the waveguide \cite{yu2014nanowire}. The overall throughput efficiency from input fiber through the device with the APCW to output fiber is approximately $20$\% for the experiments described here. 
 
The silicon chip containing a set of APCWs is affixed to a ceramic holder using heat-cured, low out-gassing glue. Each of eight devices is connected using separate input and output fibers aligned to the respective waveguide by V-grooves etched into the substrate \cite{yu2014nanowire}. The same heat-cured glue is utilized to fix the fiber position within the V-groove. This entire assembly with $16$ coupling fibers for $8$ devices is mounted in a vacuum chamber approximately $20$ mm from a magento-optical trap (MOT) from which cold cesium atoms are transported to an individual APCW by way of an optical conveyor belt described in the next section.

\section{Periodic Atom Delivery and Probe Measurements}

Previous experiments with strong coupling of atoms to nanophotonic structures have relied upon tight focusing of far-off-resonance traps (FORTs \cite{grimm2000optical}) to confine a single atom \cite{thompson2013coupling,tiecke2014nanophotonic} or several atoms \cite{gobanprl,hood2016atom} $100-200$ nm above nanophotonic cavities and waveguides. To further investigate atomic motion and atom-field interactions near nanophotonic devices, our current approach uses a moving optical lattice that repetitively delivers atoms to the surfaces and central vacuum gap of a nanoscopic APCW. Although each lattice period results in of order one atom entering the central vacuum gap of the APCW, a recursive scheme for transferring atoms from the moving lattice to stationary FORT sites within the APCW could create a $1$D lattice of single atoms with high fractional filling in time $\simeq10^{-3}$s (e.g., a sequence of $2000$ lattice periods each lasting $\simeq1$ $\mu$s), as will be discussed in the next section. Each atom thereby trapped within a unit cell would couple to single GM photons within the waveguide with interaction rate $\Gamma_{\text{1D}}$ approximately $10\times$ larger than previous schemes with similar devices \cite{yu2014nanowire,gobanprl,hood2016atom}.

The moving optical lattice of trapped atoms is prepared by loading a 1D dipole trap from atoms confined in a MOT using polarization gradient cooling (PGC) \cite{kuppens2000loading}. The dipole trap is formed using two counter-propagating $\text{TEM}_{00}$ laser beams $(B_1,B_2)$, each with waist $w_0 \simeq 60$ $\mu$m at the device and red detuning from the Cs D2 line by $\Delta_{\text{lattice}} \simeq 2\pi \times 800 $ GHz.  The atoms are thus trapped in a series of `pancakes' around intensity maxima along $z$ with spacing at $\lambda_{\text{laser}}/2$.  For each experimental trial, $\simeq2\times10^6$ atoms are loaded into the free-space lattice and the spatial extent of the atomic sample spans $\simeq1800$ pancakes. After the atoms are loaded into the 1D lattice, polarized perpendicular to the waveguide (linearly polarized along $y$), we utilize degenerate Raman sideband cooling (DRSC) to cool the atomic sample to $T\simeq12$ $\mu$K \cite{VuleticDRSC}. 

Beams $(B_1,B_2)$ are derived from a single laser beam that is split into two paths each with a double-pass 80MHz acousto-optic modulator (AOM). A frequency chirping sequence is sent to one of the AOMs to create a moving optical lattice \cite{schrader2001optical}, which acts as an optical conveyor belt to deliver atoms to the APCW at a speed given by the lattice spacing $\lambda_{\text{laser}}/2$ and the chirp frequency $f_{\text{chirp}}$, $v_{\text{lattice}}= f_{\text{chirp}}\times\lambda_{\text{laser}}/2$ .  For the experiments discussed here, the frequency difference between $B_1$ and $B_2$ is ramped from zero up to $f_{\text{chirp}} = 1.2$ MHz corresponding to an atom delivery speed of $v_{\text{lattice}}=0.51 $ m/s. The relevant diagram for loading and delivering atoms to the APCW with this conveyor belt technique is found in SI I.

The atomic interaction with the APCW is probed using a weak resonant field of frequency $\omega_{\text{probe}}$ input to either a TE or TM GM of the waveguide.  In qualitative terms, atoms located within $100-200$ nm from the surfaces of the APCW result in additional absorption and/or dispersion for a probe field transmitted at $\omega_{\text{probe}}$ relative to the device with no atoms. By recording transmission spectra obtained by scanning  $\omega_{\text{probe}}$ around a particular free-space Cs transition at frequency $\omega_{\text{a}}$, we obtain quantitative information related to light shifts of transition frequencies and modified decay rates that atoms experience in transiting near the structure \cite{hood2016atom}. More precise understanding of atomic trajectories on scales $\lesssim 50$ nm and $\lesssim 100$ ns can be gained by utilizing the periodic arrival of atoms by the moving optical lattice supplemented by auxiliary, far-from resonance GM fields that strongly perturb internal atomic states and thus atomic motion.

\section{`Clocked' Transmission Spectra}

By utilizing the lattice delivery method illustrated in Fig \ref{fig:clock}(ai-aii), atoms arrive to the APCW trapped in a sequence of pancakes, with the repetition frequency of arrivals given by $f_{\text{chirp}}$ and periodicity $\tau_{\text{lattice}} =1/f_{\text{chirp}}$.  During the passage of any one pancake, a small fraction ($\simeq10^{-8}$) of the lattice power is scattered by device imperfections into GMs of the APCW and propagates to both ends, where it is efficiently coupled into optical fibers and then separated from the probe fields using volume Bragg gratings (VBGs) \cite{Goban2012Demonstration}. The leaked lattice light (with a power of $\simeq5$ nW) is detected by an avalanche photodiode (APD), with the resulting current observed with high signal-to-noise and contrast (i.e., oscillating from near zero to a sequence of maxima with contrast $\simeq 0.7$ and period $\tau_{\text{lattice}}$ to provide a `fringe signal'). The photocurrent is directly digitized and recorded with an FPGA, as well as processed in real-time using a threshold detector that converts the fringe signal to a single time marker for each pancake, which we call the lattice synchronous markers, defined as $\tau_i$ (i.e., the red dashed lines in Fig. \ref{fig:clock}(b) at time $i\times \tau_{\text{lattice}}$ for the $i^{\text{th}}$ pancake).  These synchronous markers provide a consistent clock to register time stamps of each (separately) detected probe photon transmitted by the APCW (i.e., green bins in Fig. \ref{fig:clock}(b)) with the movement of the lattice through the structure. The random nature of the device imperfections on both the top and bottom can alter the timing of the lattice scattering into the APCW relative to the atom arrival.  While this introduces a level of uncertainty, it manifests as a global phase offset which is consistent for all measurements on a single device, but can vary between different waveguides as discussed in SI I.

The transmitted probe counts, recorded on a single photon counting module (SPCM) and digitized, from the passage of each lattice pancake can then be offset in time by lattice number (i.e., for pancake $i$, $t_\text{{i,clock}} = t_i - \tau_i$, where $\tau_i=i\times \tau_{\text{lattice}}$) and summed over all pancakes $\{i\}$ to produce a `clocked' record as illustrated schematically in Fig. \ref{fig:clock}(c), where $t$ refers to the clocked time for a sequence of $N_p$ pancakes passing near the APCW. Histograms built from the time differences between the lattice time stamps and the probe time stamps reveal microscopic information about atom motion near the APCW. For example, Fig \ref{fig:clock}(d) displays a measured histogram that clearly evidences the phase-sensitive nature of the atom arrival for a single probe detuning matching the free-space Cs D$1$ transition $F=3 \rightarrow F'=4$.  In Fig. \ref{fig:clock} and throughout the paper, the probe transmission with  atoms is normalized to the transmission through the waveguide when no atoms are present. The use of the Cs D1 transition for probing instead of the D2 cycling transition is to avoid the tensor shifts of the D2 excited states that arise from the lattice detuning from the D2 line. For lattice period $\tau_{\text{lattice}} = 833$ ns, minimum transmission (i.e., maximum loss) is observed around the clocked time $t \simeq 450$ ns, corresponding to increased atom number near the APCW as in Fig. \ref{fig:clock}(ai). By contrast, maximum transmission (and minimum loss) is evident near $t=0,833$ ns as in Fig. \ref{fig:clock}(aii) with lattice antinodes (and hence atom number for a red detuned FORT) located away from the device. 

To gain greater insight into atomic motion and internal state shifts for atoms near the APCW (e.g., AC Stark shifts and resulting forces), we create two-dimensional clocked spectra by combining measurements as in Fig \ref{fig:clock}(d) for a sequence of probe detunings, $\Delta_{\text{p}} = \omega_{\text{probe}} - \omega_{\text{a}}$. For each value of $\Delta_{\text{p}}$, time bins of recorded probe counts over the lattice period with atoms present are normalized to the probe counts at the same detuning but absent atoms in the lattice. Fig. \ref{fig:clock}(e) provides an example of a measured two-dimensional clocked spectra for a weak TM probe beam. For each detuning $\Delta_{\text{p}}$ of the probe beam, typically $5-10$ trials of the experiment are combined, with each trial consisting of $1800$ lattice pancakes and repeated every $2$ seconds.

Cuts of the $2$D spectra along lines of varying probe detuning $\Delta_{\text{p}}$ for fixed clock times $t$ (i.e., times relative to the lattice synchronization signal) can exhibit distinct spectra resulting from spatially dependent atomic density and coupling strength of atoms to the APCW. An example of a transmission spectrum at $t_{\text{cut}} = 450$ ns is shown in Fig. \ref{fig:clock}(f). Following the discussion in the SI, we fit the measured clocked transmission spectra to an effective model that was developed in \cite{hood2016atom,AsenjoGarcia:2017bm} for a random number of atoms trapped above a 1D PCW and moving along $x$. The three fit parameters for this effective model are,
$\Gamma^{\text{eff}}_{\text{1D}}$ and $J^{\text{eff}}_{\text{1D}}$, which describe effective decay rates and frequency shifts, respectively, for $N$-atom radiative interactions with the measured GM (TE or TM), as well as $\Gamma^{\prime}$ for single-atom decay into all other modes (principally free-space modes), which is assumed to have no collective effects \cite{asenjo2017sub}. 

The coupling strengths $\{\Gamma^{\text{eff}}_{\text{1D}}(t)$, $J^{\text{eff}}_{\text{1D}}(t)\}$ and decay rate $\Gamma^{\prime}(t)$ are extracted as functions of the clocked lattice time $t$ (SI). Fig. \ref{fig:clock}g provides an example of such temporal behavior of $\{\Gamma^{\text{eff}}_{\text{1D}}(t)$, $J^{\text{eff}}_{\text{1D}}(t),\Gamma^{\prime}(t)\}$ over a lattice period. Here $\Gamma^{\text{eff}}_{\text{1D}}(t)$ changes over time reflecting the atom density and coupling to the structure in a lattice period. $J^{\text{eff}}_{\text{1D}}(t)$ remains relatively constant indicating a small contribution from atom-atom interactions, which is unsurprising given that these experiments are taken far from the dielectric bandedge where the coherent coupling term is small. This technique thereby provides otherwise inaccessible information about radiative interactions along the APCW as pancakes of atoms move through the device.

\section{Clocked Spectra with Stark GMs}

For the previously described clocked spectra, atoms experience spatial and temporal variations of AC Stark shifts $\Delta^{\text{lattice}}_{\text{AC}}(r,t)$ due to the complex structure of the optical fields of the moving optical lattice and CP forces in the vicinity of the APCW, which are considered in more detail in the next section on numerical simulation. In this section we investigate experimentally clocked transmission spectra in the presence of a far-detuned GM that produces its own spatial distribution of AC Stark shifts $\Delta^{\text{GM}}_{\text{AC}}(r,t)$ near the surfaces of the APCW, as illustrated by the TM and TE GM mode profiles shown in Fig. \ref{fig:sem}(b).  The GMs here are launched incoherently from each fiber port of the APCW in order to avoid the large vector shifts on the atomic energy levels. We achieve this incoherent sum by detuning one of GMs by 100 MHz which is a fast enough beat-note that the atomic motion is unaffected, however, not large enough compared to the overall GM detuning of 60GHz to create any appreciable difference from the point of view of the atoms (i.e., an atom interacting with a GM detuned by 60 GHz behaves similarly to one interacting with a 60.1 GHz detuned GM).

\begin{figure}[h]
\centering
\includegraphics[width=1\linewidth]{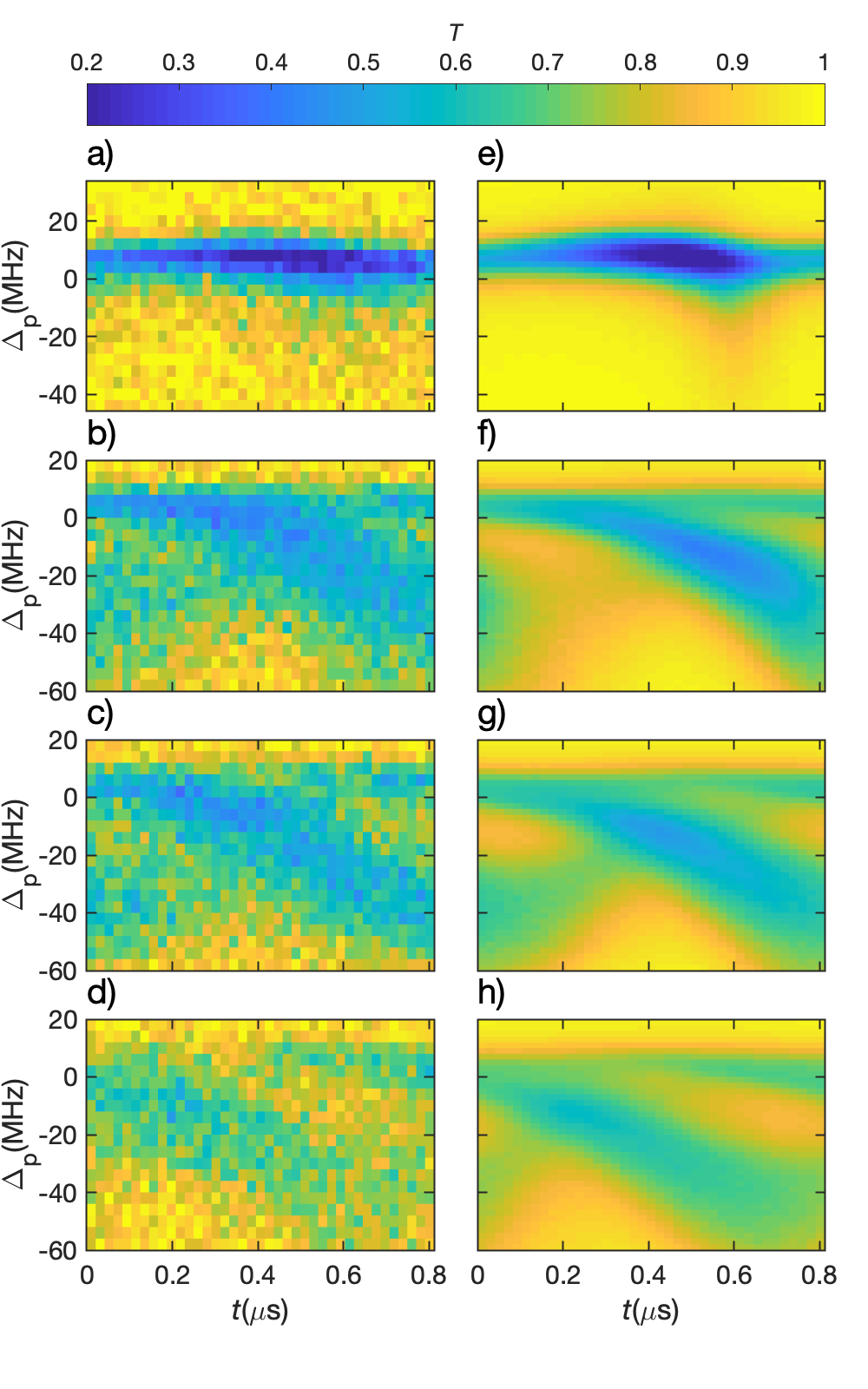}
\caption{A comparison of the data (a-d) and simulation (e-h). In both cases, there are two TM GMs of the APCW being excited, namely 1) a weak TM probe beam with variable frequency $\omega_p$ and detuning
$\Delta_p = \omega_p - \omega_a$, with $\omega_a$ the free-space frequency of the $F=3 \rightarrow F'=4$ transition in the D1 line of atomic Cs and 2) a TM Stark GM with a sequence of increasing powers from zero in the top panels. For the measurements in the left column of Fig. \ref{fig:tmtmsimdat}, the sequence of input powers is $P_{Stark}=\{0,42,66,92\}$ $\mu$W from top to bottom, while for the simulations in the right column, from top to bottom, the internal power sequence is  $P_{Stark}=\{0, 12.6, 19.8, 27.6\}$ $\mu$W for the GM detuning of $58$ GHz from $F=3$ on the D2 line. As the simulated TM Stark beam power is increased the atomic sample response alters in both time and detuning in ways similar to increasing the power in the simulation. This allows the intensity of guided modes to be calibrated by comparing the data and simulation.}\label{fig:tmtmsimdat}
\end{figure}

As shown in the left column of panels in Fig. \ref{fig:tmtmsimdat}(a-d), the addition of static GM fields leads to $2$D clocked spectra with much richer structure than absent these fields. The measured spectra exhibit variations in transmission on time scales $\delta t \simeq 100$ ns and associated length scales $\delta z \simeq 50$ nm for lattice velocity of 0.5m/s, under the influence of a TM GM with power range $0 \leqslant P_{Stark} \leqslant 92 $ $\mu$W and blue detuning of 50 GHz from the D2 line (referred to from now on as the `TM Stark GM'). The corresponding peak AC Stark shifts in the vacuum spaces surrounding the APCW are estimated to be $0 \leqslant \Delta^{\text{GM}}_{\text{AC}} \leqslant 50$ MHz, bounded by the kinetic energy of the atoms imposed by the lattice speed. Here, the atoms enter in $F=3$ and the probe beam is tuned around the $F=3\rightarrow F'=4$ transition of the D1 line of Cs (with free-space frequency $\omega_a$), again to avoid inhomogeneous broadening associated with the D2 excited-state tensor shifts from the free-space optical lattice.

In the presence of the TM Stark GM for the measurements in the left column, atoms experience a repulsive dipole force and the atomic transition frequency is shifted smaller than that of free-space. Atoms arriving to the side of the APCW facing the incoming moving lattice begin to climb up the repulsive potential created by the blue TM Stark GM and lose kinetic energy. As atoms climb this potential, the AC Stark shift they experience increases due to the spatial intensity of the TM Stark GM.  Evident in the data is that atom arrivals at different times exhibit different Stark shifts.

Moving to higher power for the TM Stark GM in the left column of Fig. \ref{fig:tmtmsimdat}, we observe an evolution of atomic signatures where atoms arriving at different times within one pancake experience different Stark shifts. A temporally varying atomic signature on the scale of $\simeq 100$ ns, while itself of interest, can be used to determine the arrival time of atoms and their spatial distribution around the surfaces of the APCW.  This information cannot be extracted from the data alone, but when aided by numerical simulations and validations, can enable `actionable' information to be obtained from the measured clocked spectra. In the next section we describe numerical simulations of atom delivery and clocked spectra, with the good correspondence between experiment and simulation already displayed in the two columns of Fig. \ref{fig:tmtmsimdat}. With this basis, we then turn to applications of these techniques.

In principle we can choose any GM polarizations for the probe and Stark fields, as well as detuning and Stark shift depending on the effect desired on the atomic signal. However, the blue TM Stark beam yields the most interesting and potentially useful results. Clocked spectra for additional GM configurations (e.g., TE probe and TM Stark beam) can be found in SI III. Additionally, here we focus on lattice speeds of $0.51$ m/s; however, we can easily adjust this speed by changing the beat frequency between $B_1$ and $B_2$, and have experimentally investigated lattice speeds from $0.7$ m/s down to $0.02$ m/s.

\section{\label{sec:level1}Simulations of Atom Motion and Clocked Spectra}

In this section, we present numerical simulations of atom trajectories and APCW transmission spectra. Numerical simulations of atom motion in a moving optical lattice and through the near fields of the APCW can be utilized to understand the nanoscopic dynamics of atoms near the APCW. Together with a model based upon optical transfer matrices for atoms located near the APCW, theoretical simulations for clocked transmission spectra can be generated for comparison to experiment, as in the right column of Fig. \ref{fig:tmtmsimdat}. Such simulations can also aid the design and operational validation of GM optical traps and test numerically techniques for loading small volume ($(\Delta x, \Delta y, \Delta z) = (30, 100, 140)$ nm) GM traps in short times ($\simeq1$ $\mu$s).

The simulations for atomic transport presented here are carried out in the 2D-space of $y,z$ to reduce the required computational resources and enable more rapid explorations of parameter space. Justification for this reduction from full 3D is that the lattice fields along $x$ for full 3D simulations exhibit only small modulation ($\lesssim 5\%$), as is also the case for so-called side-illumination (SI) traps employed in Refs. \cite{gobanprl,hood2016atom}. Full simulations in 3D are currently in development.

For the current 2D simulations, the domain area extends $\pm 25$ $\mu$m along $y$ and from $-10$ $\mu$m to  $60$ $\mu$m along $z$ in a plane that contains the thick part of the APCW (i.e., the red dashed line in Fig. \ref{fig:sem}(a)), with the center of the gap at the origin. The total time dependent potential $U(t)$ for a single atom consists of three contributions: 1) $U_{\text{lattice}}(t)$ for the optical lattice that differs significantly from the usual case of free-space due to forward and backward scattering from the APCW for each otherwise independent, counter-propagating lattice beam, with $U_{\text{lattice}}(t)$ simulated with finite element method (FEM) implemented in COMSOL \cite{Comsol53}, 2) $U_{\text{GM}}(t)$ from guided-mode fields externally input to the APCW, with $U_{\text{GM}}(t)$ determined from eigenmode calculations for the APCW GMs done with MPB \cite{Johnson:01}, and 3). $U_{\text{CP}}$ from CP force originating from the interaction between the atoms and the dielectric surfaces \cite{intravaia2011fluctuation}. Atoms are initialized in $U_{\text{lattice}}$ with a Boltzmann distribution for temperatures ranging over $10 $ $\mu$K $< T_{\text{initial}}< 150$ $\mu$K and for lattice depths $200 $ $\mu$K $< U^{\text{initial}}_{\text{lattice}}< 500 $ $\mu$K  at distance $60$ $\mu$m from the APCW in $z$, where the scattered fields from the APCW are small. Atom trajectories are calculated by solving the classical equations of motion for the assumed independent atoms in the optical dipole potential $U(r,t)=U_{\text{lattice}}(r,t)+U_{\text{GM}}(r)+U_{\text{CP}}(r)$, which consists of the optical lattice, APCW GMs and CP force. The CP force is important to include in these simulations and Fig. \ref{fig:cp} provides a comparison graph consisting of Fig. \ref{fig:tmtmsimdat} (a) and (e), and the simulation case where $U_{\text{CP}}(r)$ is removed.  The result clearly indicates that, to achieve good agreement between simulation and data, CP force must be included in the overall potential felt by the atoms.

\begin{figure*}[t]
\centering
\includegraphics[width=1\textwidth]{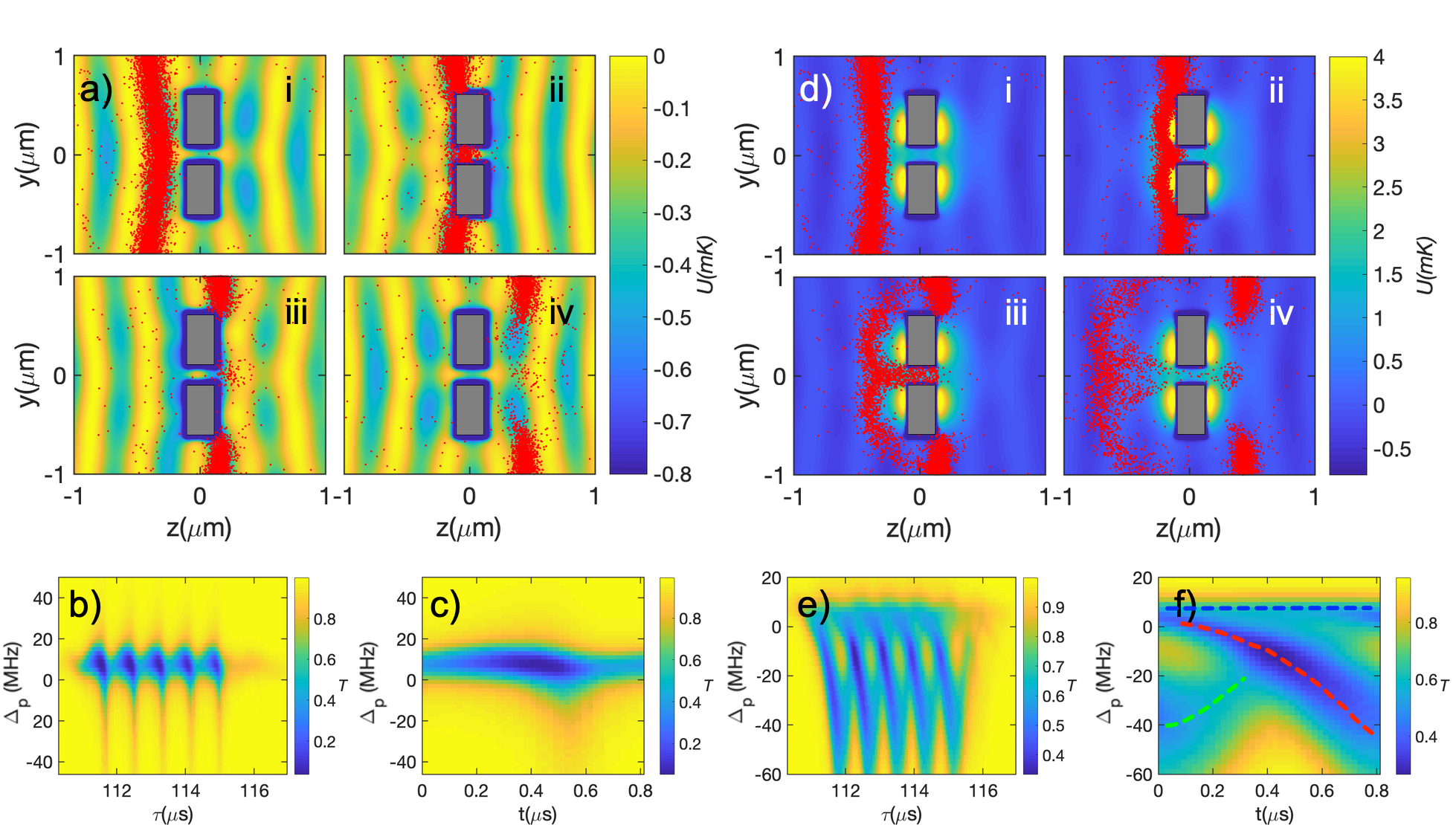}
\caption{Simulated atom delivery and the corresponding simulated clocked spectra. Note that the atom number in panels (a) and (d) is much larger per pancake than in the experiment in order to illustrate the multitude of trajectories a single atom could potentially follow. a) The atomic trajectories without any APCW GM at 4 different times. The two gray rectangles are a cross section at the thick part of the APCW, as indicated by the red dashed line in Fig. \ref{fig:sem}(a). i) shows atomic trajectories at $\tau = \tau_i$, ii) at $\tau = \tau_i + \frac{2}{3} \tau_{\text{lattice}}$, iii) at $\tau = \tau_{i+1} + \frac{1}{3} \tau_{\text{lattice}}$ (where $\tau_{i+1}=\tau_i+\tau_{\text{lattice}}$) and iv) at $\tau = \tau_{i+2}$. b) By simulating multiple pancakes and sampling from the trajectories, a spectrum of multiple pancakes can be calculated. c) A clocked spectrum generated by folding the multiple pancake spectrum. Here the lattice configuration for clocked time $t=0$ is shown in a)i and iv (i.e. the timing when the highest intensity of the lattice is at $z=0$). d) The atomic trajectories with an intense blue TM Stark beam. Timing is the same as (a). e) The transmission spectrum for 5 pancakes with a blue TM guided mode. f) Clocked spectrum with a blue TM Stark beam. Notable features of (d),(e) and (f) include: 1) Atoms that go around the sides of the APCW experience little AC stark shift from the GM and remain trapped in the optical lattice standing wave. These atoms contribute to the horizontal strip around $8$ MHz, as indicated by the blue dashed line in (f). 2) Atoms that interact with the APCW GM strongly on the surface facing incoming atoms and in the vacuum gap of the APCW. These atoms contribute to the negative AC Stark shifted feature. As atoms climb up the strong repulsive potential, the negative AC stark shift increases, creating the downward sloping feature, as indicated by the red dashed line. And as the atoms bounce back or pass through the gap, the AC Stark shift decreases in magnitude, as indicated by the green dashed line.}\label{fig:simulation}
\end{figure*}

The simulated atomic trajectories near the APCW for a single pancake with initial loading temperature $T_{\text{initial}}=100$ $\mu$K and no GM ( $U_{\text{GM}}=0$) are shown in Fig. \ref{fig:simulation}a. Notable features include 1) phase advances and retardations as each  `pancake'  nears and departs from the central plane of the APCW, which lead to atomic acceleration and deceleration, and 2) a significant flux of atoms enter the central $250$ nm vacuum gap of the APCW.

To calculate the APCW transmission spectrum as a function of time, the atom trajectories are randomly sampled according to the experimentally measured density of $\simeq 500$ atoms per pancake. The sampled atom trajectories are then distributed along the $x$ direction with probability proportional to the probe intensity in the APCW. For example, for a TE probe with frequency near the APCW TE band edge, the atom trajectories are distributed with a $\cos^2(2\pi x/a)$ probability distribution, where $a$ is the APCW unit cell spacing ($370 $ nm), to approximate the high contrast Bloch mode. For a TM probe with frequency near the Cs D1 or D2 transitions, the TM band edges are both far from the probe frequency \cite{hood2016atom} with low contrast Bloch modes (i.e., effectively traveling waves), so that the atom trajectories are distributed uniformly along $x$. Since our GM probe field is far below saturation, the transmission of the system as a function of probe frequency can be calculated with the transfer matrix model \cite{hood2016atom,AsenjoGarcia:2017bm,chang2012cavity}.

For an initial loading of five consecutive pancakes, Fig. \ref{fig:simulation}(b) shows the transmission spectrum of a weak TM probe $T(\Delta_p, t)$ calculated from the resulting simulated trajectories by way of a matrix transfer model. By applying the same analysis as previously described for our experiments (namely, folding the five pancakes spectra into one optical lattice period), a clocked spectrum can be generated over the time scale $\tau_{\text{lattice}}$, with the result shown in Fig. \ref{fig:simulation}(c). The clocked spectrum is also present in Fig. \ref{fig:tmtmsimdat}(e.)

 \begin{figure*}[t]
\centering
\includegraphics[width=1.0\linewidth]{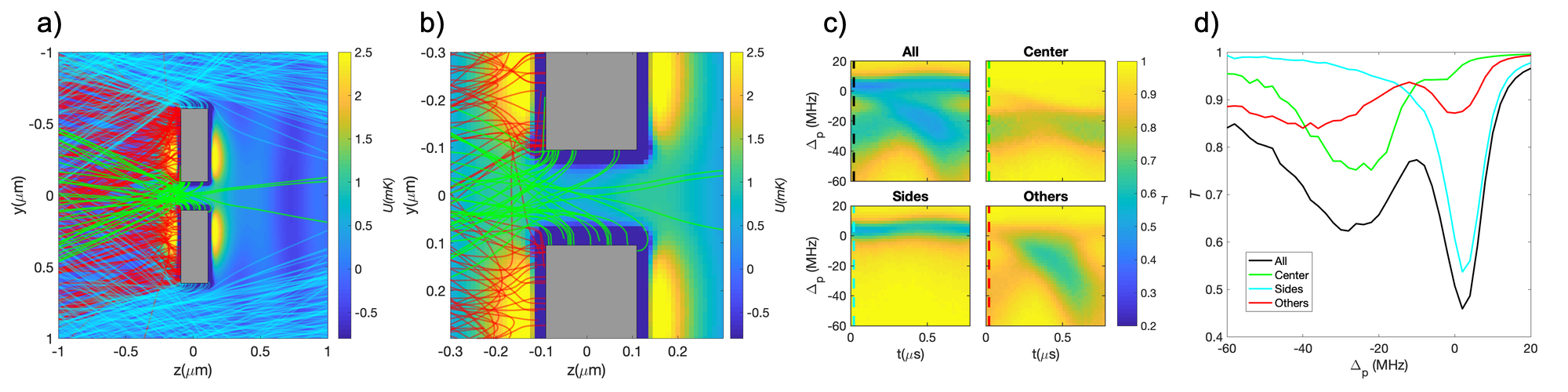}
\caption{a) Different classes of atom trajectories with a TM blue detuned Stark beam. The green trajectories are the `center' trajectories that enter the gap of the APCW. The blue trajectories are the `sides' trajectories that go around the sides of the APCW. And the red trajectories are `other' trajectories that bounce from or crash on the surface facing incoming atoms of the APCW. b) A zoomed view of the center gap showing atom trajectories entering the APCW. A large fraction of atoms penetrate deep into the gap before crashing into the dielectric wall via CP forces. c) The clocked spectra for a TE probe of the different classes of atom trajectories shown in (a) and (b). Notice the separation of different classes in both time and AC stark shift. d) A cut at $20$ ns in the clocked spectrum, as shown by the dashed lines in (c). The black line shows the spectrum of all trajectories. The color lines show the spectra of corresponding classes of trajectories in (a,b). The spectrum for all atoms (black curve) has a large peak around probe detuning $\Delta_p \simeq-20$ MHz, which predominantly results from atoms entering the central gap. For these times and detunings, another GM could be triggered in real time to selectively target gap atoms (e.g., for trapping) while they transit the vacuum gap of the APCW in successive lattice `pancakes'.}\label{fig:TMTE} 
\end{figure*}

The simulation results with a calculated blue detuned TM stark GM are shown in Figs. \ref{fig:simulation}d-f, note that the TM Stark GM alters the atom trajectories and induces AC Stark shift in the transmission spectra. With Fig. \ref{fig:simulation}(f), two classes of atoms trajectories can be identified: 1) Atoms that go around the sides of the APCW, that experience little AC stark shift from the TM Stark GM and remain trapped in the optical lattice standing wave. These atoms contribute to the horizontal strip around $8$ MHz in Fig. \ref{fig:simulation}(f), as indicated by the blue dashed line. 2) Atoms that interact with the APCW GM strongly on the surface facing the incoming atoms and in the vacuum gap of the APCW. These atoms contribute to the negative AC Stark shifted feature in Fig. \ref{fig:simulation}(f). As atoms climb up the strong repulsive potential, the negative AC Stark shift increases, creating the downward slopping feature, indicated by the red dashed line in Fig. \ref{fig:simulation}(f). For atoms that bounce back or pass through the gap, the AC Stark shift decreases in magnitude, as indicated by the green dashed line in Fig. \ref{fig:simulation}(f). More insight of atom motion around the APCW can be extract from the clocked spectrum as explained in the next section.

A side-by-side comparison of the simulation and data is shown in Fig \ref{fig:tmtmsimdat}.  This not only demonstrates reasonable correspondence between simulation and data, indicating reliability of conclusions drawn from simulations, but also provides the ability to determine the actual intensity within the APCW, a vitally important piece of information when determining the optimum power for GM trap beams.

A more detailed description of the simulations can be found in the SI along with links to movies of the atom delivery to the waveguide. The panels in Fig. \ref{fig:simulation}(a) and Fig. \ref{fig:simulation}(d) are snap shots at different times from these movies. Also, we have explored the possibility of utilizing a blue detuned lattice where the atoms are repelled from the regions of high intensity.  Markedly observed in such simulation movies, for a blue lattice, is the increased delivery of atoms to the center vacuum gap of the APCW when compared to the traditional red detuned lattice.

\section{Applications}

Fig. \ref{fig:tmtmsimdat} establishes strong correspondence between our measurements and simulations for various features of clocked spectra involving TM-probe and TM-Stark beams. Beyond the results in Fig. \ref{fig:tmtmsimdat}, we have made similar comparisons of clocked spectra for other pairs of GMs (e.g., (TE probe, TM Stark), (TM probe, TE Stark)), as described in Fig. \ref{fig:TMTE} and the SI. These results provide further validation of our numerical simulations, which are based upon calculated atomic trajectories as atoms brought by the moving lattice pass around and through nanoscopic regions of the APCW in the presence of forces from the lattice itself, Stark GMs, and surface forces. With a reasonable degree of confidence, we can then attempt to use the measured clocked spectra to provide insight into atomic trajectories that are beyond direct observation in our current setup.

An example of the general intuition is as follows. As illustrated in Fig. \ref{fig:sem}(b), a TM mode is primarily sensitive to atomic trajectories that intersect the top of the waveguide, while for the TE mode, the regions of highest intensity are on the sides and in the gap between the two beams. Hence, a TE probe can improve our sensitivity to gap and side atoms. In order to distinguish between these two classes of trajectories (side and gap), a TM Stark GM with blue detuning (as in Fig. \ref{fig:TMTE}(a,b)) can be used to separate atoms interacting with the TE probe into two distinguishable classes. Because of the AC-Stark shift and the spatial intensity distribution of the TM blue GM, an atom entering the vacuum the gap of the APCW will experience a larger AC Stark shift than an atom passing around the side. By validating this effect in our simulations, we can then utilize it to separate experimentally the side and gap classes of atoms in a clocked spectrum.

One important application is to employ the lattice delivery method to achieve high factional filling of the trap sites within the APCW by way of a recursive loading scheme (i.e., some small probability to transfer one atom from the moving lattice to a GM trap for the passage of each successive lattice `pancake'). Such recursive loading requires detailed understanding of the experimental signatures of atoms entering the central vacuum gap of the APCW, including the probability with which atoms actually enter the central gap. With the reliability of the simulations validated, we have some confidence in the result in Fig. \ref{fig:TMTE}(a) and Fig. \ref{fig:TMTE}(b), which shows that a significant fraction of atoms delivered to the APCW do indeed enter the vacuum gap under appropriate conditions of lattice speed and intensity for a TM Stark GM.

This being the case, the next step is to identify the operational signatures for the class of `gap' atoms in a clocked spectrum, which we have done in Fig. \ref{fig:TMTE}(c) by way of a TE probe GM. Here, the total clocked spectrum based upon all atom trajectories from the simulation is decomposed into a set of clocked spectra for individual trajectory classes, including the class of trajectories that pass through the central vacuum gap. Figure \ref{fig:TMTE}(d) zooms into the clocked spectrum by examining spectral cuts at successive time intervals to identify times and detunings for which the gap atoms have distinct spectral signatures. Clearly, the spectrum for all atoms (black curve) has a large peak around probe detuning $\Delta_p \simeq-20$ MHz, which predominantly results from atoms entering the central gap at clocked times around $t \simeq 20$ ns. For these times and detunings, another GM could be triggered in real time to selectively target gap atoms while they transit the vacuum gap of the APCW in successive lattice `pancakes'.

Inspired by Ref. \cite{bannerman2009single} for `single-photon cooling', we are attempting to adapt the capabilities illustrated in Fig. \ref{fig:TMTE} for the transfer of atoms from the moving lattice into a GM trap by way of a single cycle of optical pumping (i.e., a single-photon scattering event)\cite{taieb1994cooling,falkenau2011continuous}. As depicted in the SI, consider a moving lattice with FORT potential $U_{\text{lattice}}$ and a TE-GM trap $U_{\text{GM}}(F)$ within the APCW having the property that $U_{\text{GM}}(F_1)\gg U_{\text{lattice}}$ for the ground manifold level $F_1$ and $U_{\text{GM}}(F_2) \lesssim U_{\text{lattice}}$ for level $F_2$, which for definiteness could be the $6S_{1/2}, F_1=3,F_2=4$ levels in Cesium. Assume that atoms being transported by a lattice with velocity $v_{\text{lattice}}$ such that their kinetic energy is much less than $U_{\text{GM}}(F_1)$ and are prepared in $F_2$.

Then by employing techniques similar to those illustrated in Fig. \ref{fig:TMTE}, it should be possible to identify clocked time $t$ and detuning $\Delta_{\text{OP}}$ at which to trigger a pulse of TE guided-mode light for optically pumping atoms $F_1 \rightarrow F_2$ within the central vacuum gap. Since the transfer involves momentum $\leqslant 2 \hbar k$ and $U_{\text{GM}}(F_1)\gg U_{\text{lattice}}$, atoms should be trapped within $U_{\text{GM}}(F_1)$ as the lattice minimum exits the trap region. Furthermore, the choice of detuning, $\Delta_{\text{OP}}$, for the optical pumping pulse allows discrimination in favor of atoms near the coincident minima of $U_{\text{GM}}(F_2)$ and $U_{\text{GM}}(F_1)$ (i.e., near the center of a (random) unit cell of the APCW) by way of AC-Stark shifts as in Fig. \ref{fig:TMTE}.

Beyond this discussion of single-photon trapping, rather more practical applications of clocked spectra are already in place in our laboratory. Examples include 1) the use of clocked spectra for intensity calibration within a GM of the APCW resulting from a known input power at the optical fiber coupled to the sequence of waveguides that lead to the actual APCW\cite{yu2014nanowire} and 2) validation of trap parameters for GM traps formed from the summation of red and blue GMs \cite{Goban2012Demonstration,hung2013trapped}.

Certainly, the agreement between simulation and measurement can be considerably improved by extending the simulations to more faithfully capture the complexity of the APCW, including fully 3D lattice transport and vector fields of the APCW, and implementing more accurate calculations of CP potentials for all surfaces of the APCW as in Ref. \cite{hung2013trapped}, which is nontrivial. Such improvements could enable more advanced measurements to be undertaken, such as quantitative validations of CP potentials in the spirit of Refs. \cite{cronin2009optics,bender2014probing}. 

\begin{figure}
\centering
\includegraphics[width=1\linewidth]{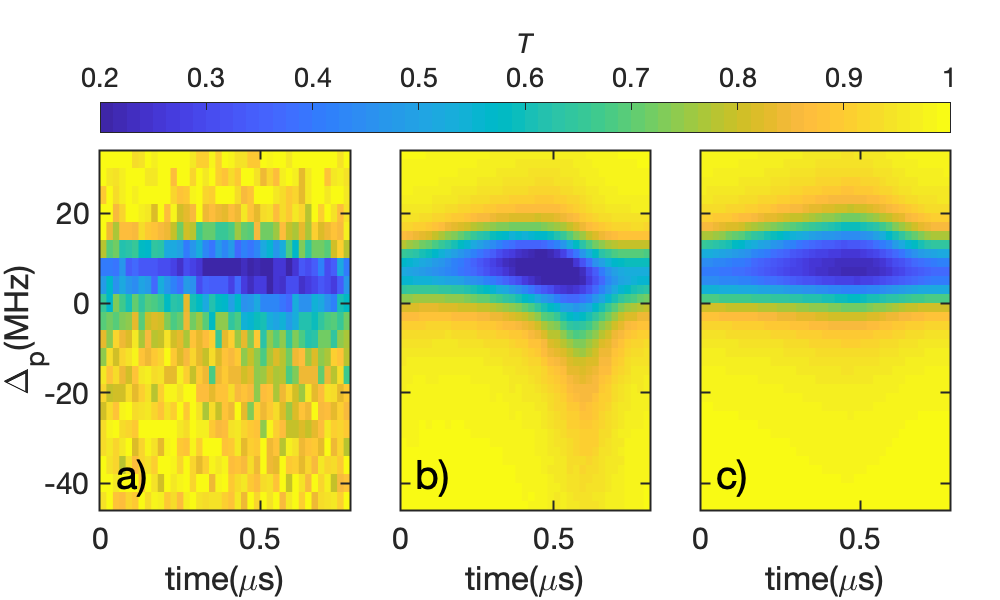}
\caption{A comparison of clocked spectra for a) experimental data, b) simulation with the CP potential and c) simulation without the CP potential, for the case of no Stark GM and probing with a weak TM probe. The addition of the CP potential is essential for obtaining good correspondence between experiment and simulation.}\label{fig:cp}
\end{figure}

A full calculation of Casimir-Polder (CP) potential of an atom around the non-trivial geometry of our dielectric structure is extremely involved and require huge numerical resources. To simplify the calculation, the $C_3$ coefficient of short range van der Walls atom-surface attraction for both Cs ground-state and $6P_{1/2}$ excited state are taken from Ref. \cite{Failache2003}. The CP potential for a single nano-beam in the APCW is approximated by only considering the atom-surface attraction force of the atom and its closest dielectric surface. A full calculation of the CP potential as in Ref. \cite{hung2013trapped,gobanprl} of the APCW is under development. However, as seen in Fig. \ref{fig:cp}, our current measurements are already sensitive to the $C_3$ coefficient chosen for the Cs $D1$ transition in our simplified van der Waals model \cite{cronin2009optics,gregoire2016analysis}.  This simple theory, though not complete, is nonetheless important in generating good correspondence between experiment and simulation.

A quite different application of our system of clocked-atoms delivered to a PCW is utilization of this time-dependent, optically dense atomic medium for novel nonlinear optical experiments, such as soliton propagation, as investigated in Ref. \cite{kozhekin1995self}. The relevant pulse durations are much smaller than the atomic lifetime $\tau \simeq 30$ ns, which is in turn much shorter than the atom transit time through the APCW, so the pulses could be triggered to interact with an optical system with selected values of $ \Gamma_{1\text{D}}^{\text{eff}},J_{1\text{D}}^{\text{eff}},\Delta_{\text{AC}}, \Gamma^{\prime}$ as in Fig. \ref{fig:clock}(g) by varying the offset of input pulse and lattice clock timing.

In this non-exhaustive discussion of applications, we finally mention the possibility of opto-mechanics. In our modeling and experiments, the GM probe is kept weak to avoid affecting atomic motion as well as saturation, allowing modeling of optical transmission and reflection by way of matrix transfer techniques. That is, atoms strongly affect the probe response, but the probe only weakly affects atomic motion. On the other hand, the Stark GMs in the work presented are far detuned so that their transmission is only weakly modified by the atoms, but clearly atomic motion is strongly influenced by sufficiently intense Stark GMs, as in Fig. \ref{fig:simulation}. Consider now that the probe and Stark GMs are one and the same field with intensity sufficient to drive atomic saturation as well as to create optical forces that modify atomic motion. We would then be in a regime requiring a self-consistent description of the internal and external degrees of freedom for atoms and optical fields within the APCW (i.e., a nonlinear regime for opto-mechanics with atoms). Although we have carried out such measurements with our current experiment, the results are beyond our existing simulation capabilities.

\section{Conclusion and Outlook}

Trapping and cooling atoms at distance $\simeq100$ nm from the internal surfaces of dielectric photonic crystal waveguides (PCWs) requires a new set of advances relative to standard atomic physics techniques. Towards this end, we have described experiments and numerical simulations to understand better the motion of cold atoms under the influence of forces from GM fields within a PCW, external illumination with light transverse to the PCW axis, and CP interactions at the surfaces of the PCW. In particular, a moving optical lattice has been utilized to transport cold atoms to the surfaces and vacuum spaces of a nanoscopic PCW. By way of synchronous detection of the passing optical lattice and EM transmission spectra, we find rich phenomenology related to temporal and spatial variations of AC-Stark shifts, radiative absorption from and emission into GMs, and (vacuum) surface forces, all of which strongly affect atomic motion. 

In addition to empirical characterization of the resulting phenomena, we have carried out extensive numerical simulations in an attempt to achieve effective modeling tools with quantitative prediction capabilities, as well as provide physical insight. Initial validation of the simulations has been made by way of direct comparisons of simulated and measured clocked transmission spectra and reasonable correspondence achieved. By way of our simulations, we can then infer underlying characteristics of atomic motion and internal fields, and to some modest degree, control atomic dynamics within and near the PCW (e.g., maximizing the number of atoms that pass through the central vacuum gap of the PCW and providing operational signatures manifest in the associated clocked spectra).

Understanding the interplay of forces from GM traps, moving optical lattices, and surfaces of the APCW on the motion of atoms constitutes a significant step towards GM trapping of arrays of atoms within unit cells along the PCW. Observations of single and collective radiative phenomena would then become possible in engineered photonic environments, as described theoretically in Refs. \cite{gonzalez2015subwavelength,douglas2015quantum,hung2016quantum,chang2018colloquium}. 

Throughout this manuscript, we have treated the atoms as classical point particles without interactions. Surely, at densities of $10^{12}$ atoms/cm$^3$, encountered around and within the APCW, these assumptions are no longer completely valid and a richer set of atomic interactions are accessible. Future experiments should investigate more carefully density dependent effects, including inelastic hyperfine changing collisions and light-assisted collisions, which are critical for loading free-space optical tweezers \cite{schlosser2001sub,kaufmanprx}.

Moreover, we have ignored the wave character of the atoms (e.g., at $v_{\text{lattice}} = 0.5$ m/s the deBroglie wavelength $\lambda_{\text{dB}} \simeq 6$ nm). However, as the atomic velocity is reduced during cooling and trapping, $\lambda_{\text{dB}}$ can become comparable to the dimensions of the vacuum spaces in the APCW (e.g., at $10$ $\mu$K, $v_{\text{atom}} = 0.03$ m/s, and $\lambda_{\text{dB}} \simeq 100$ nm). In this regime, novel physics might be manifest as we arc back to the beginning of atom interactions with periodic nano-structures \cite{keith1988diffraction,cronin2009optics}. 

Aside from trapping atoms in these nano-structures one can imagine specifically designed experiments tailored toward more precise measurements of CP forces at dielectrics surfaces. The ubiquity of the lattice delivery method does not require a structure as complicated as the APCW and in fact a more simple structure would ease the transition of our 2D simulations to a full 3-dimensional simulation model (a step we are currently implementing for our  APCW structure).

The Authors acknowledge sustained and important interactions with  A. Asenjo-Garcia, J.B. B\'eguin, D.E. Chang and his group, A. Goban, J. D. Hood, C.-L. Hung, J. Lee, X. Luan, Z. Qin, and S.-P. Yu. We carried of the nanofabrication in the Caltech KNI and the clean room of O.J. Painter whom we gratefully acknowledge.  HJK acknowledges funding from the Office of Naval Research (ONR) Grant \#N00014-16-1-2399, the ONR MURI Quantum Opto-Mechanics with Atoms and Nanostructured Diamond Grant \#N00014-15-1-2761, the Air Force Office of Scientific Research MURI Photonic Quantum Matter Grant \#FA9550-16-1-0323, the National Science Foundation (NSF) Grant \#PHY-1205729, and the NSF Institute for Quantum Information and Matter Grant \#PHY-1125565.

\bibliography{main}
\end{document}


\preprint{APS/123-QED}

\title{Supplemental Information \\ Clocked Atom Delivery to a Photonic Crystal Waveguide}

\author{A. P. Burgers}
\email{aburgers@caltech.edu}
\author{L. S. Peng}
\author{J. A. Muniz}
 \altaffiliation[Present address: ]{JILA, University of Colorado, Boulder, CO 80309.}
\author{A. C. McClung}%
 \altaffiliation[Present address: ]{Department of Electrical and Computer Engineering, University of Massachusetts Amherst, 151 Holdsworth Way, Amherst, MA 01003.}
\author{M. J. Martin}%
 \altaffiliation[Present address: ]{P-21, Physics Division, Los Alamos National Laboratory, Los Alamos, NM 87545.}
\author{H. J. Kimble}
 \email{hjkimble@caltech.edu}
\affiliation{Norman Bridge Laboratory of Physics MC12-33, California Institute of Technology, Pasadena, CA 91125.}
\maketitle


\date{\today}

\section{Experimental Set-up}
The experimental process starts by pushing, with a near resonant pulsed beam, a continuously loaded MOT in the source chamber down a differential pumping tube to the science chamber \cite{metcalf2012laser}.  The pushed atoms are recaptured in the science chamber MOT where we load them into a 1D lattice from two counter-propagating beams originated from the same laser, $B_1$ and $B_2$. Both beams are passed through acousto-optic modulators (AOMs) to control the relative frequency between $B_1$ and $B_2$ along with intensity and to achieve fast switching times. The 1D lattice loads from atoms in the science MOT by way of an interval of polarization gradient cooling (PGC), which is followed by a degenerate Raman sideband cooling (DRSC) interval to achieve a final axial temperature of 12 $\mu$K \cite{metcalf2012laser}. 

Once the atomic sample has been loaded and cooled, one of the AOMs is frequency chirped from the normal operation at 80 MHz to 81.2 MHz (for the case of the experimental data presented, however, as stated in the manuscript, the final chirp difference frequency can be arbitrarily defined). This chirp sequence is achieved by utilizing a signal generator set to 70 MHz and mixing with a 10MHz signal from a programmable direct digital synthesizer (DDS)(Analog Devices AD8954). A tunable filter is used to block the lower frequency component from mixing. The chirping sequence is written to the DDS memory using an `Arduino DUE' microcontroller. The chirping sequence ramps the frequency from 10MHz to the desired RF frequency (here 11.2 MHz to achieve a 1.2MHz beat and a lattice speed of 0.51 m/s), thereby creating a moving optical lattice. The atoms are conveyed over a distance of $20$ mm to the center region of a particular APCW device via the moving $1D$ optical lattice (i.e., `optical conveyor belt') with temperature in the moving lattice frame (typically $\sim 10-30$ $\mu$K) much less then the lattice depth (typically $\sim 300-500$ $\mu$K).

As the confined atomic cloud passes a chosen APCW, atoms near the waveguide are interrogated by a weak guided-mode (GM) probe injected into the APCW with frequency $\omega_p$ tuned around the atomic free-space resonance $\omega_a$.  The transmitted probe beam is separated spectrally from lattice light that scatters into the APCW by a volume Bragg grating (VBG), as well as from light in any other guided modes (GMs), which are used in some experiments. The transmitted and reflected probe light is detected by single-photon counting modules (SPCMs) with a time stamp recorded for each detected photon. Scattered lattice light that emerges in a GM is likewise detected, both by a SPCM, as well as an analog APD to produce time-stamps and a real-time zero crossing signal, respectively, which is likewise recorded to fix sequential lattice periods. The lattice time tags and the probe time tags are registered to each other and a `clocked' histogram created for a single lattice period as described in the manuscript (Fig. 3).  
 
One experimental concern is determining the relative phase between the atomic signature and the lattice sync signal. Due to the (assumed) random distribution of scatterers on the surfaces of the waveguide, the time of largest scattering into the waveguide can vary relative to the time of maximum intensity centered in the gap of the APCW (i.e., what we define as $t=0$ for the simulated clocked spectra). Since the thickness of the device is 200 nm with a refractive index $n_{SiN}=2$, the optical distance is 400 nm, which is comparable to the distance between adjacent pancakes ($\sim425$ nm) in free-space. This can cause a shift the clocked spectrum by roughly half the lattice period $\pm\tau_{\text{lattice}}/2$ depending on the distribution of scatterers for each different device surface. We have modeled such processes by calculations of the field intensity of the moving lattice in various regions of the APCW (e.g., $10$ nm depth of the front or back surface, and the inner and outer walls) and numerically found offsets $\simeq \pm\tau_{\text{lattice}}/2$. Furthermore, the scattered power into the waveguide from each individual lattice beams $B_1$ (incident on the front surface) and $B_2$ (incident on the back surface) is generally not equal, though the powers are the same in free-space, where this ratio varies from device to device. This supports the supposition that scattering of lattice light into GMs arises from fabrication-dependent defects and not systematic imperfections.
\begin{figure*}
\centering
\includegraphics[width=\textwidth]{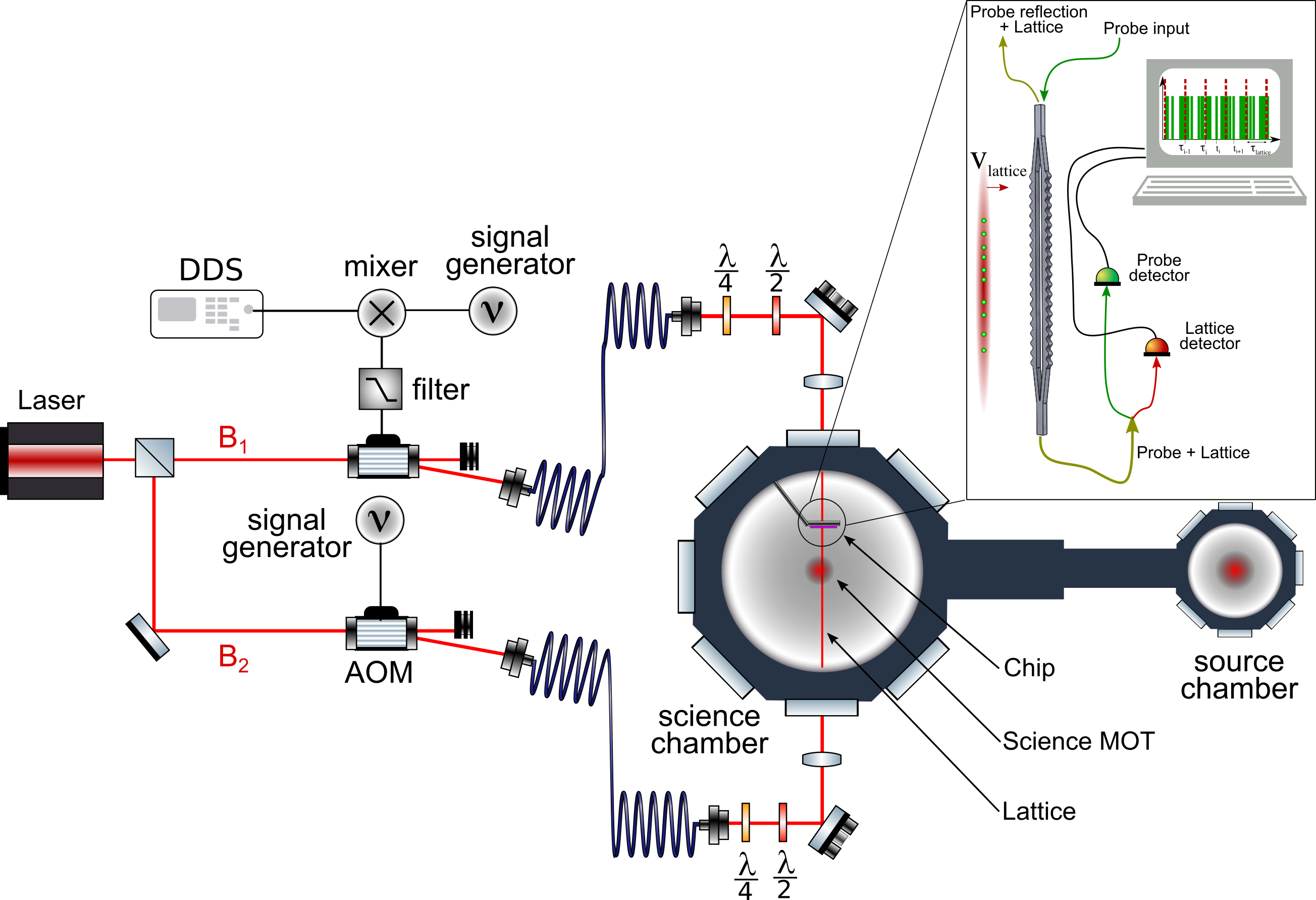}
\caption{Schematic of the experimental apparatus.  The Science chamber is loaded by pushing a source MOT from the Source chamber through a differential pumping tube to the Science chamber. The 1D optical lattice (conveyor belt) originates from a TiSaph laser and is split sending each beam, $B_2$ and $B_2$, to an acousto-optic modulator (AOM).  One of the AOMs receives a frequency chirp sequence from a direct digital synthesizer (DDS) creating a moving optical lattice that conveys the atoms to a particular APCW on a chip with multiple such devices. (Inset) The atoms interact with the waveguide through GMs of the structure, in particular a weak probe tunned around the atomic free-space resonance.  The probe light and lattice light are separated by Volume Bragg Gratings (VBGs) and each detected and digitized to create the histograms presented in the manuscript.}
\end{figure*}

Operationally, we use two methods to correct the measured clock spectra to account for time offsets between the lattice sync signal and the `true' time for which lattice maxima from successive lattice periods cross the center line of the APCW along $z$. First of all, we employ a method for which a clocked spectrum with no GM Stark beam is summed over all detunings $\Delta_p$ for a given laboratory offset time (i.e., project a clocked spectrum onto the time axis). The minimum optical depth  in time is then offset in time to correspond to clocked time $t=0$. 
A second method utilizes a cross-correlation technique between the probe counts at a single detuning and the lattice fringe signal by offsetting the relative timing in such a way that the highest correlation point is at lattice time $t=0$. Both these methods agree within the $100$ ns uncertainty of our timing logic for the lattice sync signal. 

We emphasize that the offset found for a given device determined by these techniques is constant but varies from device to device over the expected range $\simeq \tau_{\text{lattice}}/2$. For more detailed comparisons between a set of measured and simulated spectra as displayed in Figure 5, we allow a small additional offset for the data set as a whole. 

\section{Fitting $\Gamma^{\text{eff}}_{\text{1D}}$ and $J^{\text{eff}}_{\text{1D}}$ from Clocked Data} 

We fit the measured `clocked' spectra to a transmission model of the probe field through the PCW that was developed in \cite{hood2016atom,AsenjoGarcia:2017bm}, which is expressed as follows:

\begin{align}\label{eqn:t}
T(\Delta_p,t)=\left|\frac{\Delta_p+i\Gamma'/2}{(\Delta_p+J^{\text{eff}}_{\text{1D}}(t))+i(\Gamma'+\Gamma^{\text{eff}}_{\text{1D}}(t))/2}\right|^2.
\end{align}
Here $\Delta_p$ is the detuning between the probe frequency and the free-space atomic resonance frequency, $\Gamma'$ is the atomic decay rate into all modes (mostly free-space) other than the GM of interest, and $\Gamma^{\text{eff}}_{\text{1D}}(t)$ and $J^{\text{eff}}_{\text{1D}}(t)$ are the emission rate into the waveguide and the atom-atom coupling rate for the GM of interest, respectively. The time dependence of $\Gamma^{\text{eff}}_{\text{1D}}$ and $J^{\text{eff}}_{\text{1D}}$ arises from the periodic arrival and transit of the atoms. For the ideal case of a single lattice `pancake' with atomic and probe frequencies near the band edges of the APCW, we would find that 
\begin{align}
\Gamma^{\text{eff}}_{\text{1D}}(t)=\sum_{i=1}^{N_{\text{at}}}\Gamma^{ii}_{\text{1D}}\left(r_i(t)\right),
\end{align}
where $N_{\text{at}}$ is the number of atoms within a single pancake and $r_i$ is the position of the $i^{\text{th}}$ atom.
 
However, in our experiment it is difficult to disentangle the number of atoms interacting with the waveguide and the spatial variation of $\Gamma_{\text{1D}}$ for different atoms and trajectories. Hence, we introduce $\Gamma^{\text{eff}}_{\text{1D}}$ as an effective atomic coupling to the waveguide, and likewise for the term $J^{\text{eff}}_{\text{1D}}$, which was found to be quantitatively adequate for the analyses in \cite{hood2016atom}. 
In microscopic terms,  $J^{ij}_{\text{1D}}$ and $\Gamma^{ij}_{\text{1D}}$ relate to the real and imaginary components of the Green's function for radiative interactions between atoms $(i,j)$, mediated by the GM of interest of the APCW.
 
From the effective model of Eq. \ref{eqn:t}, the coupling strengths $\Gamma^{\text{eff}}_{\text{1D}}(t)$ and $J^{\text{eff}}_{\text{1D}}(t)$ can be extracted as functions of the clocked lattice time by taking detuning cuts of the 2D spectrum at fixed times in Fig. 3(e) and fitting each spectrum to the above model.  An example of measurements at one time slice and the corresponding fit of the spectrum is shown in Fig. 3(f). Such fits can be performed for each time $t$ in a 2D clocked spectrum to extract the temporal behavior of the fit parameters versus $t$ and thereby obtain further information about atomic couplings to the waveguide on a microscopic scale as the successive lattice periods of atoms move through the device.

\section{Clocked Spectra with Different Guided Modes}
\begin{figure}
\centering
\includegraphics[width=1\linewidth]{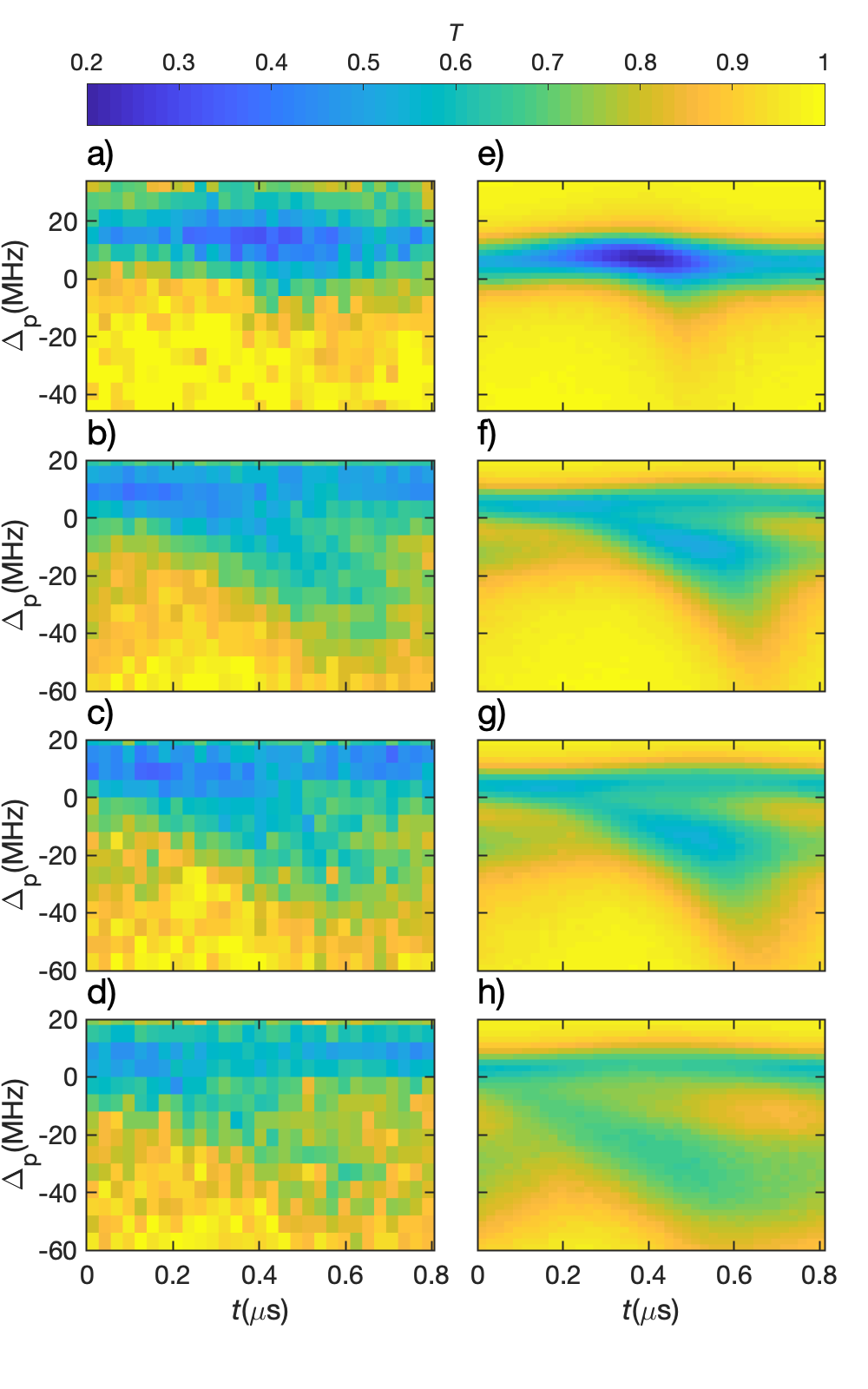}
\caption{A comparison of data (a-d) and simulation (e-h). A blue detuned TM Stark GM is excited in the APCW, and the atoms are probed with a weak TE probe with detuning $\Delta_p$. For the measurements in the left column of the Figure, the sequence of input powers is $P_{\text{Stark}}=\{0,20,30,52\}\mu$W from top to bottom. For the simulations in the right column, from top to bottom, the internal power sequence is  $P_{Stark}=\{0, 3.7, 5.6 , 9.7\}\mu$W for the GM detuning of $58$ GHz from $F=3$ on the D2 line. 
\label{fig:blueTMTE}}
\end{figure}

Here we present additional experiment and simulation results. Fig. \ref{fig:blueTMTE} shows the measured and simulated spectra for a blue-detuned TM Stark GM with TE probe.  Evident in this figure is that the agreement between simulation and experiment, while still good, is not quite so good compared to Fig. 4 in the main text. We suspect this is due to the usage of TE GMs for probing, which have a complicated modulation pattern in the $x$ direction that is not addressed in the current 2D simulations \cite{hood2016atom}.  We are working towards implementing 3D simulations to better characterize the trajectories in the presence of the full APCW. This TE probe configuration is utilized to gain sensitivity to the atomic trajectories that enter into the APCW vacuum gap, as indicated in Fig. 6 from the main text.  From Fig. 6 in the main text, vacuum gap atoms appear separated from the other trajectories (`side' and `others') in detuning at $\sim 0.16$ $\mu$s. The clocked spectra in Fig. \ref{fig:blueTMTE}, provide preliminary evidence for the atoms entering the vacuum gap, which is essential for loading any FORT within the APCW.

\section{Simulations}
\begin{figure}[h]
\centering
\includegraphics[width=0.7\linewidth]{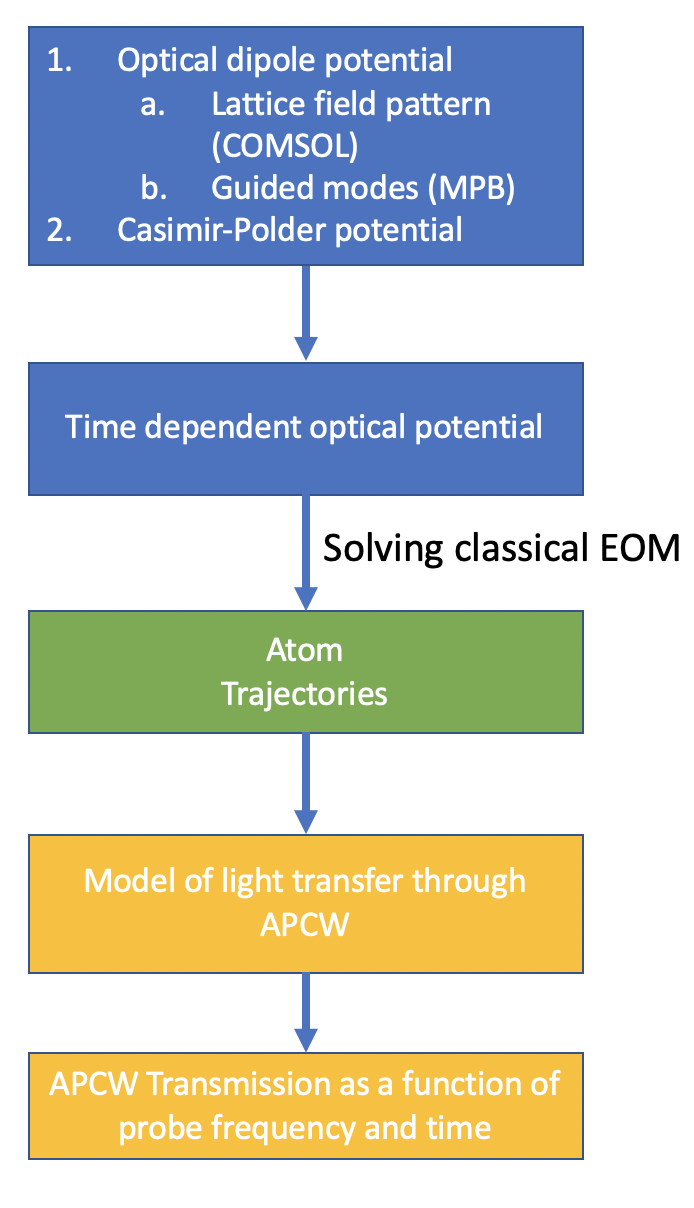}
\caption{The flowchart of the numerical simulation. First, the optical dipole potentials and the Casimir-Polder potential are calculated and combined to form the total time-dependent potential. Then, atoms are initialized into the optical potential far from the APCW and the equations of motion solved to generate the atom trajectories. The transmission $T(\Delta_p,t)$ as a function of probe detuning and time is then calculated with the transfer matrix model.}
\label{fig:flowchart}
\end{figure}

Here we present the details of the numerical simulations of atom trajectories and APCW transmission spectra. First, 2D atom trajectories $(y(t), z(t))$ are calculated by solving the equations of motion of atoms in optical dipole and Casimir-Polder potentials. Then, the atom trajectories are sampled and distributed along the $x$ direction with the probability distribution of $P(x)$, depending on the probe intensity profile (i.e. TE or TM mode). The APCW transmission spectra can be calculated with the transfer matrix model by representing `distributed' atom ${i}$ at position $(x_i(t),y_i(t),z_i(t))$ and the waveguide segment between atoms with transfer matrices. The simulation process is summarized in the flowchart in Fig. \ref{fig:flowchart}.
 
The simulations for atomic transport are carried out in the 2D-space of $y$,$z$ to reduce the required computational resources and enable more rapid explorations of parameter space. Justification for this reduction from full 3D is that the lattice fields along $x$ for full 3D simulations exhibit only small modulation ($\lesssim 5\%$), as is also the case for so-called side-illumination (SI) traps employed in Refs.\cite{gobanprl,hood2016atom}. Full simulations in 3D are currently in development. For the 2D simulations in this work, the center gap of the APCW is located at the origin, the simulation domain extends $\pm 25$ $\mu$m along $y$ and from $-10$ $\mu$m to $60$ $\mu$m along $z$ in a plane that contains the thick part of the APCW, with atoms initialized at $z \approx 60$ $\mu$K moving toward $-z$ direction.
 
The total potential $U(t)$ for an assumed independent atom consists of three contributions: 1). The optical lattice, $U_{\text{lattice}}(t)$, which differs significantly from free-space due to forward and backward scattering from each, otherwise independent, counter-propagating lattice beam off the APCW.  Here $U_{\text{lattice}}(t)$ is calculated using the finite element method (FEM) implemented in COMSOL \cite{Comsol53}. 2).  GM fields, $U_{\text{GM}}(t)$, input to the APCW, where $U_{\text{GM}}(t)$ is determined from eigenmode calculations for the APCW GMs done with MPB \cite{Johnson:01}. 3). Finally, the Casimir-Polder potential,  $U_{\text{CP}}$, originating from the interaction between the atoms and the dielectric surfaces \cite{intravaia2011fluctuation}. Potentials 1) and 2) are the ground-state optical dipole potential, which can be calculated from the field intensity \cite{grimm2000optical}. Atoms are initialized in $U_{\text{lattice}}$ with a Boltzmann distribution for temperatures ranging over $10$ $\mu$K $ < T_{\text{initial}}< 150$ $\mu$K and for lattice depths $200$ $\mu$K $ < U^{\text{initial}}_{\text{lattice}}< 500$ $\mu$K at a distance $60$ $\mu$m from the APCW in $z$, where the scattered fields from the APCW are small. Atomic trajectories are calculated by solving the classical equations of motion with SUNDIAL differential equation solver \cite{hindmarsh2005sundials} for the assumed independent atoms in the potential $U (r,t)=U_{\text{lattice}}(r,t)+U_{\text{GM}}(r)+U_{\text{CP}}(r)$. For the current parameters for our experiments, the nonadditive corrections of optical potential and CP potential \cite{PhysRevLett.121.083603,fuchs2018casimir} is estimated to be small for all the beams involved.

\subsection{Transfer Matrix Model}
\begin{figure}
\centering
\includegraphics[width=1\linewidth]{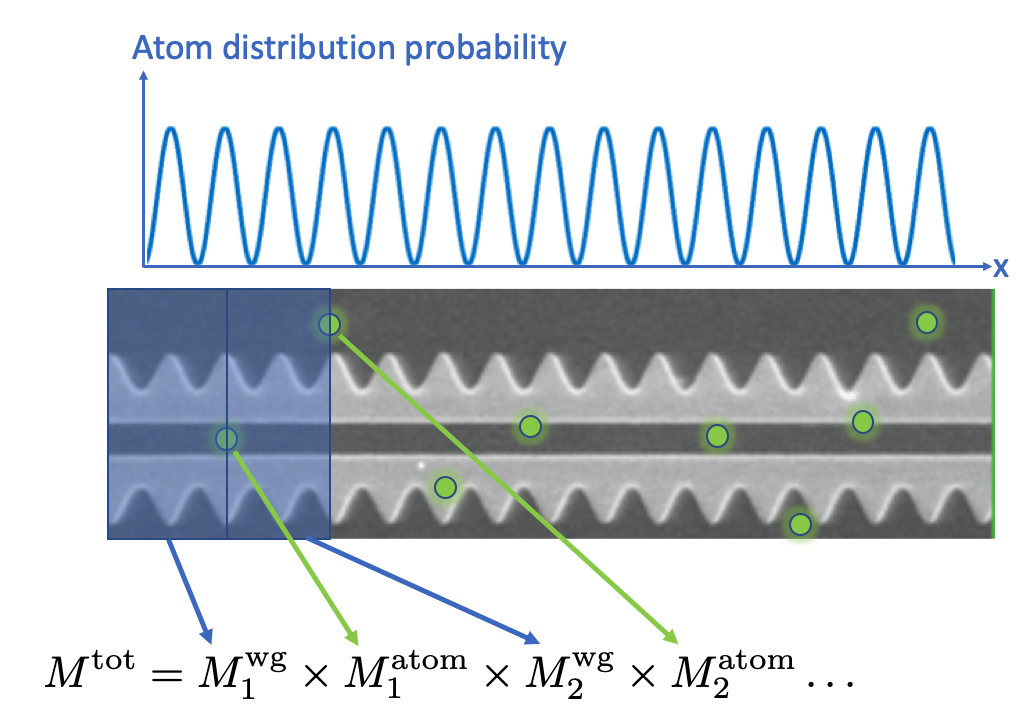}
\caption{Diagrammatic look at the transfer matrix model.  For a TE probe simulation, atoms are distributed along the APCW in the $x$ direction weighted by the sinusoidal intensity distribution of a light at the TE dielectric band edge. Light propagation along the APCW is modeled with the transfer matrix model. Each atom and the waveguide segments between adjacent pairs of atoms are represented with a transfer matrix, with the total transfer matrix being the product of all transfer matrices. The transmission of the whole system can be extract from the total transfer matrix.}
\label{fig:transfermatrix}
\end{figure}

To calculate the APCW transmission spectrum as a function of time, the atom trajectories are randomly sampled according to the experimentally measured density of $\simeq 500$ atoms per pancake. The sampled atom trajectories are then distributed along the $x$ direction with probability proportional to the probe intensity in the APCW, as shown in Fig. \ref{fig:transfermatrix}. For example, for a TE probe with frequency near the APCW TE band edge, the atom trajectories are distributed with a $\cos^2(2\pi x/a)$ probability distribution, where $a$ is the APCW unit cell spacing ($370$ nm), to approximate the high contrast TE Bloch mode. For a TM probe with frequency near the Cs D1 or D2 transitions, the TM band edges are both far from the probe frequency \cite{hood2016atom} with low contrast Bloch modes (i.e., effectively traveling waves), so that the atom trajectories are distributed uniformly along $x$. Since our GM probe field is far below saturation, the transmission of the system as a function of probe frequency can be calculated with the transfer matrix model \cite{hood2016atom,AsenjoGarcia:2017bm,chang2012cavity}.\\
Light propagation along the APCW and atoms system can be modeled with the transfer matrix model. For a probe with detuning $\Delta_p$ relative to shifted ground-state and excited-state transition frequency, the transfer matrix of an atom is:
\begin{equation}
M^{\text{atom}}(\Delta_p, \Gamma_{\text{1D}}, \Gamma^{\prime}) = 
 \begin{pmatrix}
  t-\frac{r^2}{t} & \frac{r}{t}\\
  -\frac{r}{t} & \frac{1}{t}\\
 \end{pmatrix}\label{eqn:matom}
 \end{equation}
 where $r = -\frac{\Gamma_{\text{1D}}}{\Gamma_{\text{1D}}+\Gamma^{\prime}-i2\Delta_p }$ and $t=1+r$ \cite{chang2012cavity}.\\
 And for a waveguide of length $l$ and angular wavenumber $k$,
 \begin{equation}
 M^{\text{wg}}(k,l) = 
 \begin{pmatrix}
 e^{ikl}&0\\
 0&e^{-ikl}\\
 \end{pmatrix}\label{eqn:mwg}
 \end{equation}
 with the atom trajectories distributed along the $x$ direction $(x(t),y(t),z(t))$.
To calculate the transmission at time $t$, the $i^{\text{th}}$ atom can be modeled with the transfer matrix $M^{\text{atom}}_i(\Delta, \Gamma_{\text{1D}}, \Gamma^{\prime})$ in Eq. \ref{eqn:matom}, with the detuning $\Delta_p$ calculated from ground-state and excited-state light-shifts induced by the lattice, GM and CP potential. The emission rate into the waveguide $\Gamma_{\text{1D}}$ is proportional to the probe intensity profile, and $\Gamma^{\prime}$ is the decay rate into free-space and other GMs. The waveguide segment $i$ between atom $i-1$ and atom $i$, can be modeled with eqn. \ref{eqn:mwg}. The total transfer matrix $M^{\text{tot}}$ is the product of all transfer matrices along the waveguide, $M^{\text{tot}} = \prod_{i = 1}^{n} (M^{\text{wg}}_i\times M^{\text{atom}}_i) $, where $n$ is the number of the sampled atoms, as shown in Fig. \ref{fig:transfermatrix}. The transmission of the APCW and atoms system can then be extract from the total transfer matrix $M^{\text{tot}}$.

\section{Guided Mode Traps}
\begin{figure}[h]
\centering
\includegraphics[width=1\linewidth]{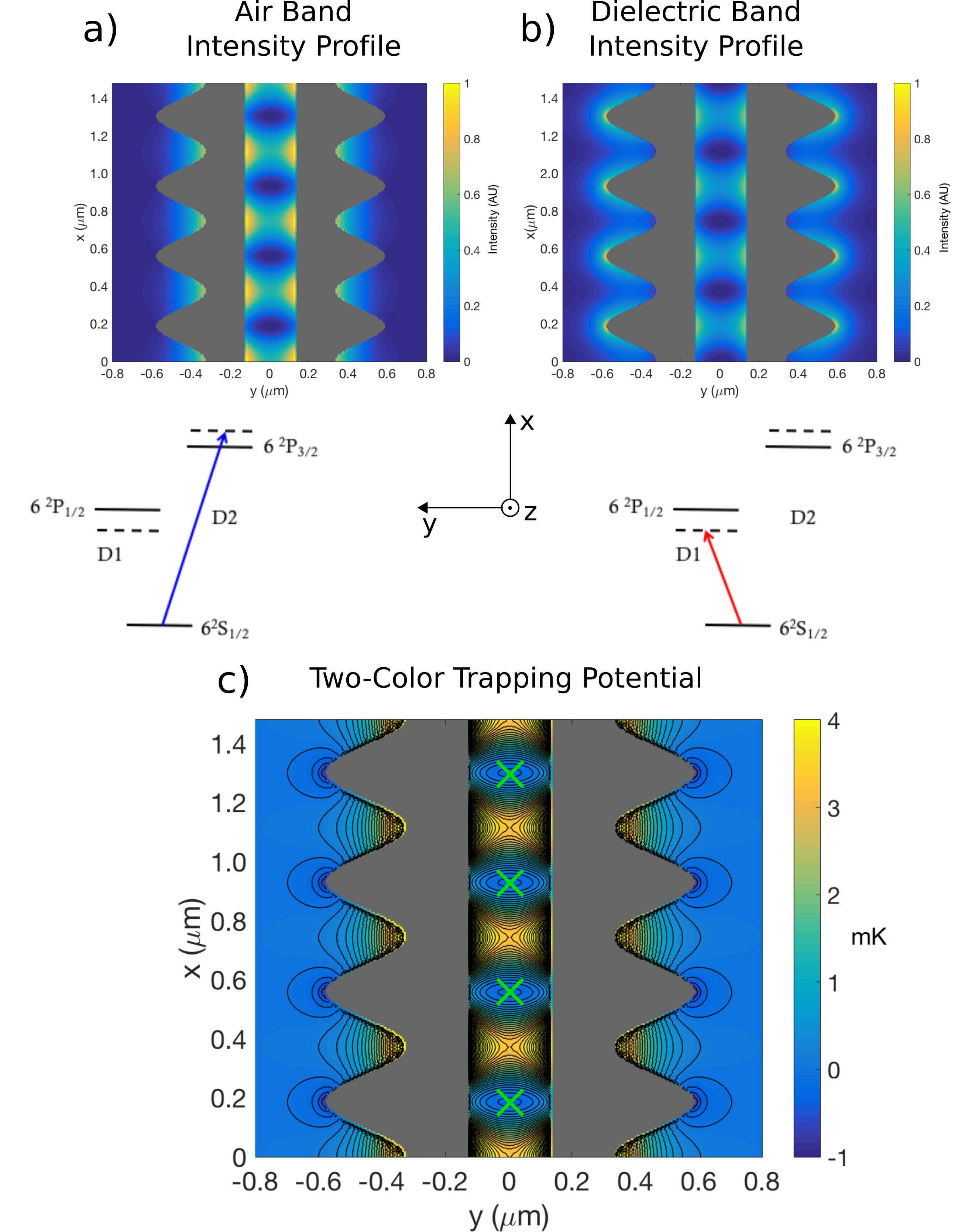}
\caption{Stable trap sites within the alligator photonic crystal waveguide (APCW) created using blue and red detuned beams at the air bandedge and dielectric bandedge \cite{hung2013trapped,yu2014nanowire}. a) By way of a repulsive optical dipole force, atoms are kept off the walls of the structure. b) An attractive optical dipole force creates a periodic trapping potential along the length ($x$ direction) of the waveguide.  This red detuned trap light also creates an attractive potential primarily utilized to create confinement out of the page ($z$ direction).  c) Total optical trap with the green `X's' indicating stable trapping points.}\label{fig:twocolor}
\end{figure}

The guided modes (GMs) of the alligator photonic crystal waveguide (APCW) can be utilized to create stable trapping potentials in the vacuum gap between the dielectric beams \cite{hung2013trapped,yu2014nanowire}. GMs at each band edge exhibit a periodic structure within the APCW as seen in Fig. \ref{fig:twocolor}.
A two color trap utilizes the periodic structure of the APCW by tuning one GM to a higher frequency (60 GHz) than the free-space atomic resonance so that the dipole force repels the atoms from this band edge.  The second GM is tunned to a frequency lower than the atomic resonance (600 GHz) creating an attractive potential.  Though these dipole force GMs are at different wavelengths, the Bloch modes of the structure set the scale for the trap locations. The result of this two-color trap scheme leads to the trapping potentials shown in in Fig. \ref{fig:twocolor}.

\begin{figure}
\centering
\includegraphics[width=0.8\linewidth]{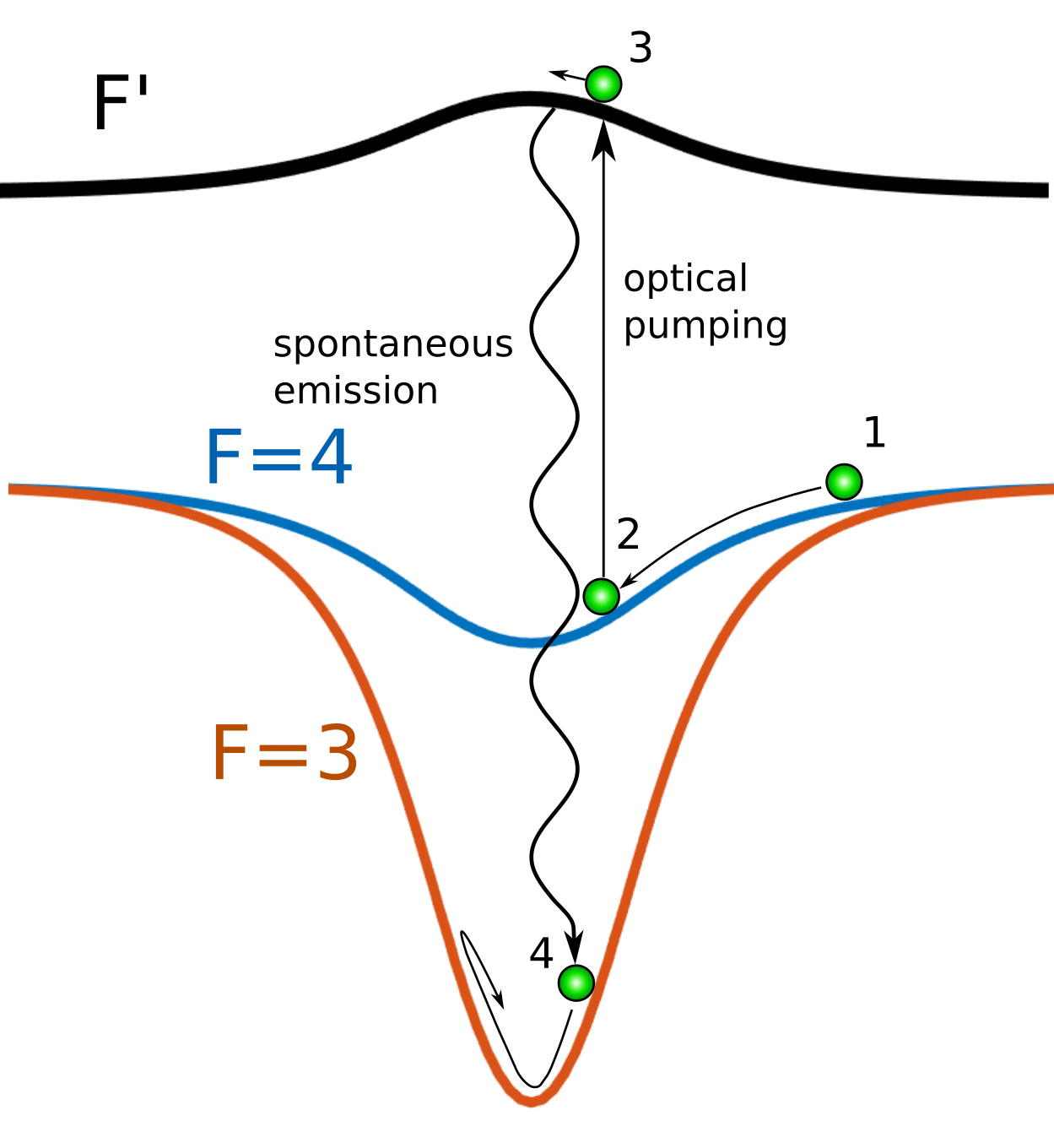}
\caption{A schematic of the implementation of the scheme in Ref. \cite{bannerman2009single} here illustrated for a cross-section of the GM trap. Atoms arrive into the trap within the APCW structure in F=4 and accelerate down the F=4 potential (point 1) . When the atom is near the center of a unit cell of the APCW, an optical pumping pulse is triggered exciting the atom to a higher electronic F' state (point 2).  The atom is in the excited state for a short time before decaying through spontaneous emission (point 3). If the atom decays to F=3 it retains approximately the same kinetic energy but now the potential barrier around it is larger than the atom's kinetic energy (point 4). The atom is thus trapped on the F=3 trap surface.}\label{fig:stateTrap}
\end{figure}

For the data presented in our manuscript, atoms are traveling at $0.51$ m/s through these trapping potentials, transiting the roughly $100$ nm trap size in only $200$ ns. To achieve trapping, the atoms must cooled and trapped in times shorter than traditional mechanisms for laser cooling and trapping of atoms. Hence we must incorporate a faster trapping scheme, here based upon that utilized in Ref. \cite{bannerman2009single}.  The atoms arrive in  one of the ground state manifolds, say F=4, and enter the trap region.  The trap is configured in such a way that the trap surface for F=4 is shallow compared to F=3, so that the atoms pick up little additional kinetic energy as they move down into the trap. When the atoms are near the trap center an optical pumping pulse promotes population to an electronic excited state where it decays with roughly equal probability to F=3 or F=4.  Atoms decaying to F=4 will simply continue their motion and `roll' out of the conservative potential; however, atoms decaying to F=3 will retain approximately the same kinetic energy only now on the different trap surface for F=3. The kinetic energy of the atoms is now insufficient to overcome the trap potential for F=3 and the atoms are now trapped.  A schematic of this process is provided in Fig. \ref{fig:stateTrap}.  In the trapping scheme described above, triggering the optical pumping pulse to initiate the state transfer must be done at a specific time to ensure the atoms are transferred to near the minimum of the F=3 surface.  Clocked delivery provides important information about when the atoms are arriving into the center of the APCW, but perhaps more importantly, the clocked signal provides us with a trigger for the optical pumping pulse. This is another useful advantage to utilizing the clocked delivery method we describe in the manuscript. 
\section{Simulation Movies}
The following movies present two cases of simulated trajectories for no TM Stark GM (Movie 1) and including a TM Stark GM (Movie 2).  The potentials present in the simulation are determined by the free-space conveyor belt lattice, the CP potential near the dielectric surface (particularly visible in Movie 1 as the constant potential around the dielectric structure) and GMs of the waveguide (only present in Movie 2). The black dots represent individually calculated atomic trajectories initialized in 5 separate pancakes and launched 60 $\mu$m from the waveguide at time $\tau=0$ (time counter observed in upper left corner of the movies).  The black dots become red a single frame before the trajectory intersects the boundary of the dielectric structure and are removed from the simulation.  The lattice speed in free-space is $0.51$ m/s, lattice depth is 500 $\mu$K, and initial temperature $T=150$ $\mu$K for atoms trapped in the lattice. Note that the number of atoms is much larger per pancake for the movies (ie 20000 atoms per pancake) than the experiment ($\approx 500$ atoms per pancake) to illustrate the multitude of trajectories a single atom could potentially follow. The two gray rectangles are a cross section at the thick part of the APCW, as indicated by the red dashed line in Fig. 2(a).\\

\subsection{Movie 1: Red lattice delivery of atoms with no Stark GMs} This movie is of atomic trajectories using the conditions of Fig. 5(a) in the main text (no Stark GM).  The four frames in Fig. 5(a) are generated using a single pancake of atoms from this simulation movie. The link to the movie can be found here: \underline{\href{https://dx.doi.org/10.14291/vd6s-6h38}{https://dx.doi.org/10.14291/vd6s-6h38}}

\subsection{Movie 2: Red lattice delivery with blue detuned TM Stark GM} This movie portrays the conditions of Fig. 5(d) in the main text (in the presence of a TM Stark GM). Here, the blue detuned TM Stark GM repels atoms from the dielectric surfaces and imposes position and time dependent AC-Stark shifts. As before, the four frames in Fig. 5(d) indicate the evolution of a single pancake taken from this simulation movie. The link to the movie can be found here: \underline{\href{https://dx.doi.org/10.14291/r6xg-j678}{https://dx.doi.org/10.14291/r6xg-j678}}
\bibliography{supplement}